\documentclass[aps]{revtex4}
\usepackage{makeidx}
\usepackage{amssymb}
\usepackage{latexsym}
\usepackage{amsmath}
\usepackage{graphicx}% Include figure files
\usepackage{bm}% bold math

\newcommand{\bra}[1]{{\langle #1 |}}
\newcommand{\ket}[1]{{| #1 \rangle}}
\newcommand{\braket}[2] {{\langle #1 | #2 \rangle}}
\newcommand{\rv}{{\bf r}}
\newcommand{\vv}{{\bf v}}
\newcommand{\Rv}{{\bf R}}

\newcommand{\qv}{{\bf q}}

\newcommand{\kv}{{\bf k}}

\newcommand{\ep}{\epsilon}
\newcommand{\w}{\omega}

\newcommand{\hw}{\hbar \omega}

\newcommand{\kvG}{{\bf k}_{\Gamma}}
\newcommand{\kvX}{{\bf k}_X}

\newcommand{\fX}{f^{X}}
\newcommand{\G}{\Gamma}
\newcommand{\BG}{{\bf \Gamma}}
\newcommand{\bSA}{{\bf \Sigma}^A}
\newcommand{\bSR}{{\bf \Sigma}^R}
\newcommand{\bSl}{{\bf \Sigma}^<}

\newcommand{\grl}{g^{\lhd R}}

\newcommand{\bgr}{{\bf g}^{R}}
\newcommand{\bga}{{\bf g}^{A}}
\newcommand{\bgl}{{\bf g}^{<}}

\newcommand{\gal}{{g}^{\lhd A}}

\newcommand{\gll}{{g}^{\lhd <}}

\newcommand{\ggl}{{g}^{\lhd >}}

\newcommand{\bt}{{\bf t}}
\newcommand{\btt}{\tilde{\bf t}}

\newcommand{\bS}{{\bf S}}
\newcommand{\alp}{\alpha}
\newcommand{\bb}{\beta}
\newcommand{\Iop}{\hat{I}}

\newcommand{\bJ}{{\bf J}}
\newcommand{\ba}{{\bf a}}
\newcommand{\bA}{{\bf A}}
\newcommand{\bGl}{{\bf G}^<}
\newcommand{\bGA}{{\bf G}^A}
\newcommand{\bGR}{{\bf G}^R}
\newcommand{\GL}{\Gamma^{\cal L}}
\newcommand{\GR}{\Gamma^{\cal R}}
\newcommand{\fL}{f^{\cal L}}
\newcommand{\fR}{f^{\cal R}}
\newcommand{\fG}{f^{\Gamma}}

\begin{document}
\title{Non-Equilibrium Green Functions in Electronic Device Modeling}
\author{Roger K. Lake}
\author{Rajeev R. Pandey}
\affiliation{Department of Electrical Engineering, University of California, Riverside, CA 92521-0204\\
Ph: 951-827-2122; FAX: 951-827-2425; email: rlake@ee.ucr.edu\\
Jan. 4, 2005\\
To appear in Handbook of Semiconductor Nanostructures, American  Scientific Publishers, eds. A. A. Balandin and K. L. Wang}

\maketitle


\section{Abstract}
We present an overview of electronic device
modeling using non-equilibrium Green function techniques.
The basic approach developed in the early 1970s
has become increasingly popular during the last 10 years.
The rise in popularity was driven first
by the experimental investigations of mesoscopic
physics made possible by high quality semiconductor heterostructures
grown by molecular beam epitaxy.
The theory has continuously been adapted to address
current systems of interest moving from the mesoscopic
physics of the late 1980s to single electronics to
molecular electronics to nanoscaled FETs.
We give an overview of the varied applications.
We provide a tutorial level derivation of the
polar optical phonon self-energy \cite{LakeNemoTheory}.
Then, focusing on issues of a non-orthogonal basis
used in molecular electronics calculations,
we derive
and the basic
Green function expressions starting from their definitions
in second quantized form in a non-orthogonal basis.
We derive the equations of motion for the retarded Green
function $G^R$ and the correlation function $G^<$, and
we derive the standard expressions for the electron density
and the current that are in widespread use.
We point out common approximations and open questions of which
one finds little discussion in the literature.

\section{\label{sec:intro}Introduction}
The applications of nonequilibrium Green
functions \cite{Keldysh,KadanoffBaym}
have been extensive
including quantum optics \cite{Haug88},
quantum corrections to the Boltzmann transport
equation \cite{Jauho84,MahanArticle}, high field
transport in bulk systems \cite{Bertoncini_Jauho_PRL92},
and electron transport through nanoscaled systems.
Our interest has been in this last category of
electron transport through nanoscaled materials
under a finite applied bias.
Below, we review the work in this area and then provide
tutorial derivations of the standard expressions.

Over the last decade, non-equilibrium Green's function
(NEGF) techniques have become
widely used in corporate, engineering,
government, and academic laboratories for modeling
high-bias, quantum electron and hole transport
in wide variety of materials and devices:
III-V resonant tunnel diodes \cite{Kim_Arnold_RTD88,AndaFlores,ChenTing_ac,LakeDattaSLATTMICRO,Birman_bistability,Grein_93,ChenTing_noise,Lakebigone,rogerpower,AndaFlores_noise,Gyungock_Kim,Klimeck_APL95,LakeNanoMes96,DRC96_proceedings,DRC97_proceedings,LakeNemoTheory,BowenMultiband2},
electron waveguides \cite{Mikesthesis},
superlattices used as quantum cascade lasers \cite{Lee_Wacker02},
Si\index{Si}
tunnel diodes \cite{CR_APL1,CR_JAP1}\index{tunnel diode},
ultra-scaled Si
MOSFETs \cite{Jovanov_iwce00,Svizhenko_JAP02,Dresden02,Svizhenko_TED03,Venugopal_JAP_5_03}\index{MOSFET},
Si nanopillars \cite{Y-J_Ko,Nanotech03,Rivas_Lake},
carbon nanotubes \cite{Nardelli99,Taylor_ab_init_PRB01,Taylor_Mag_CNT_PRL00,Taylor_Res_CNT_PRB01,Taylor_C60_PRB01,Anant_contacts_APL01,Anant_PRL02,Taylor_IV_N_CNT_PRB02,Cuniberti_PhysE02,Verges_PRL03,XueRatner_SchottkyCNT03,Register_JAP04}\index{carbon nanotubes},
metal wires \cite{Brandbyge02,Louis_NEGF_PRB03},
organic molecules \cite{WTian_JChemPhys98,Seminario_NEGF01,Xue_JChemPhys01,Y_Xue_thesis,Ghosh_Datta_PRB01,Seminario_Comp_mol01,Seminario_gain_J.Am.Chem.Soc01,Taylor_rectification_PRL02,Verges_Nanotech02,Schon_PRL02,Verges_PRB03,Xue_PRB03,Seminario_ProcIEEE03,Nitzan_asymmetric03,Seminario_program.diode03,Xue_PRB04,Ghosh_gating04,Koentopp_PRB04}\index{molecules},
superconducting weak links \cite{WeakLinks94},
and magnetic leads \cite{Taylor_Mag_CNT_PRL00,Bulka_mag_leads04,Krompiewski_spin_JPCM04}.
Physics that have been included are open-system
boundaries\index{boundaries!open-system} \cite{Caroli_I},
full-bandstructure \cite{Davidovich_Anda,Gyungock_Kim,LakeNemoTheory,DRC97_proceedings,CR_APL1,CR_JAP1}\index{bandstructure!full},
bandtails\index{bandtails} \cite{CR_JAP1},
the self-consistent Hartree
potential \cite{Birman_bistability,DRC96_proceedings,klimeck_last_nemo_pub},
exchange-correlation potentials within a density functional
approach \cite{DRC96_proceedings,Seminario_NEGF01,Taylor_ab_init_PRB01,Xue_JChemPhys01,Ghosh_Datta_PRB01,Brandbyge02,Xue_ChemPhys02},
acoustic,
optical, intra-valley, inter-valley, and inter-band phonon
scattering\index{scattering!phonon}, alloy disorder\index{alloy
disorder} and interface roughness\index{interface roughness}
scattering in Born type
approximations \cite{AndaFlores,LakeDattaSLATTMICRO,Grein_93,Lakebigone,LakeNanoMes96,DRC96_proceedings,LakeNemoTheory,CR_APL1,CR_JAP1,Lee_Wacker02}\index{Born},
photon absorption and emission \cite{Lee_Wacker02},
energy\index{energy transport} and heat
transport \cite{rogerpower} \index{heat transport}, single electron
charging and nonequilibrium Kondo
systems \cite{Groshev_PRL91,ChenTing91,HershPRL91,Hersh1,Meir_Wingreen_non_x_PRL93,Wingreen_non_x,XueRatner_single.e.tunnel03},
shot noise \cite{ChenTing_noise,Hersh2,AndaFlores_noise},
A.C. \cite{ChenTing_ac,AnantAC1,AnantAC2,Wingreen_NEGF_t,Jauho_NEGF_t,Stafford_PRL96,Kral_ac96},
and transient response \cite{Wingreen_NEGF_t,Kral_Jauho_99}.
Time-dependent calculations
are described further in \cite{Haug_Jauho_book96}.
General tutorials can be found in \cite{DattaTextbook2,Datta_SLMS00}.

The general formalism for NEGF calculations of current
in devices was first described in a series of papers in the early 1970s
 \cite{Caroli_I,Caroli_II,Caroli_III,Caroli_IV}.
The partitioning of an infinite system into left contact, device,
and right contact, and the derivation of the
open boundary self-energies for a tight-binding model
was presented in \cite{Caroli_I}.
This theory was re-derived for a continuum representation in \cite{Caroli_II},
tunneling through localized impurity states was treated in \cite{Caroli_III},
and a treatment of phonon assisted tunneling was derived in \cite{Caroli_IV}.
In 1976, the formalism was first applied
to a multi-band model (2-bands)
to investigate tunneling \cite{Bandy_Glick} and
diagonal disorder \cite{Bandy_Glick2},
and in 1980 it was extended to model time-dependent
potentials \cite{Caroli_time_dep}.

The citation rate of the first article in the series \cite{Caroli_I}
gives a good indication of the use
of the NEGF formalism applied to electronic transport
over the last 3 and a half decades.
Of the 277 citations listed, 110 occur during the years
from 2000 to the present,
119 occur during the 1990s,
22 citations occur during the 1980s, and 26 citations
occur during the 1970s.
Over the last decade and a half, the
citation rate has been steadily increasing.
The motivation for the development of the NEGF tunneling
formalism was the metal-insulator-metal
tunneling experiments that received much attention during the
1960s \cite{Duke_tunneling}.
The revival, or accelerated use, of the approach
was motivated by the experimental investigations of mesoscopic
physics made possible by high quality semiconductor heterostructures
grown by molecular beam epitaxy.
In 1988, Kim and Arnold appear to be the first to apply the NEGF
formalism to such a system, specifically,
a resonant tunneling diode \cite{Kim_Arnold_RTD88}.
As experimental methods progressed allowing finer
manipulation of matter and probing into the
nanoscale regime, the importance of quantum effects
and tunneling continuously increased,
and the theory adapted to address
the current systems of interest moving from mesoscopics to
single-electronics to nano-scaled FETs to molecular electronics.

\section{One Dimensional Transport through Planar Semiconductor Devices}
\subsection{NEMO}
Effort developing a quantum semiconductor
device simulator for devices in which the
potential varies along one dimension (1D)\index{1D}
through planar semiconductors
peaked during
Texas Instruments'\index{Texas Instruments}
Nanotechnology Engineering
program which ran from 1993-1997 and resulted in the
the Nanoelectronic Engineering Modeling (NEMO)\index{NEMO}
program \cite{ISCS_Proc_NEMO_97}.
The Nanoelectronic Engineering Modeling research program was
initially conceived to model resonant tunnel diodes (RTDs)
and, specifically, to discover the processes that
determine the minimum valley current.
The program included both theoretical development and
software development. Furthermore, somewhat unique to this
program, there was a strong experimental verification effort
which consisted of the growth, fabrication, and measurement of hundreds
of resonant tunnel diodes of varying geometries, layer thicknesses,
and material systems forming a large test matrix.

\begin{figure}[h]
\begin{center}
\resizebox{3.5in}{!}{ \includegraphics{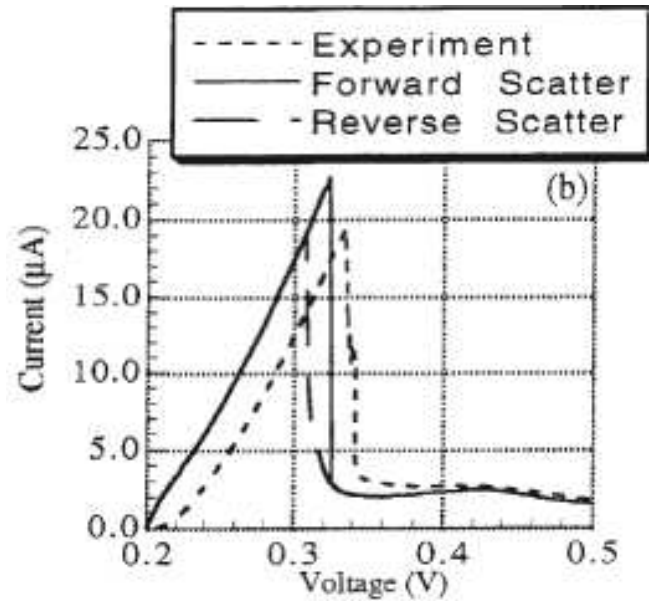} }
\end{center}
%\caption{\mbox{  }}
\caption{Example calculation including incoherent scattering from acoustic phonons, polar optical phonons and interface roughness.
The simulations are compared against experimental data.
Reprinted from \cite{DRC96_proceedings}. Copyright 1996 IEEE.}
\label{fig:drc96_phonon_peak}
\end{figure}

There were several major thrusts to the theory and modeling component:
(a) a flexible treatment of the open system
boundaries \cite{Klimeck_APL95,Lake_VLSI_Design,LakeNemoTheory},
(b) charge self-consistency \cite{DRC96_proceedings,DRC97_proceedings,klimeck_last_nemo_pub}
(c) incoherent scattering
processes \cite{LakeNanoMes96,DRC96_proceedings,LakeSST98,LakeNemoTheory,klimeck_last_nemo_pub} and
(d) full bandstructure models \cite{DRC97_proceedings}.
An example of a comparison between theory and experiment is shown in
Fig. (\ref{fig:drc96_phonon_peak}). This is a simulation of a GaAs / AlAs
RTD with a 5.66 nm well, 3.1 nm barriers, and 20 nm spacer layers at a temperature
of 4.2K. The phonon peak is clearly visible. It rides on a background current
resulting from interface roughness scattering. The scattering current is
calculated self-consistently with the Hartree potential and a
local density approximation (LDA) for the exchange-correlation.
The magnitude of the scattering assisted valley current matches well the
experimental data. It was found that the Hartree potential significantly
overestimated the intrinsic bistability \cite{ZouWillander1/94}.
Even with scattering and the LDA potential, the bistability is
overestimated since it is observed in
the simulations where it is not observed in the experiment.

\begin{figure}[h]
\begin{center}
\resizebox{6.5in}{!}{ \includegraphics{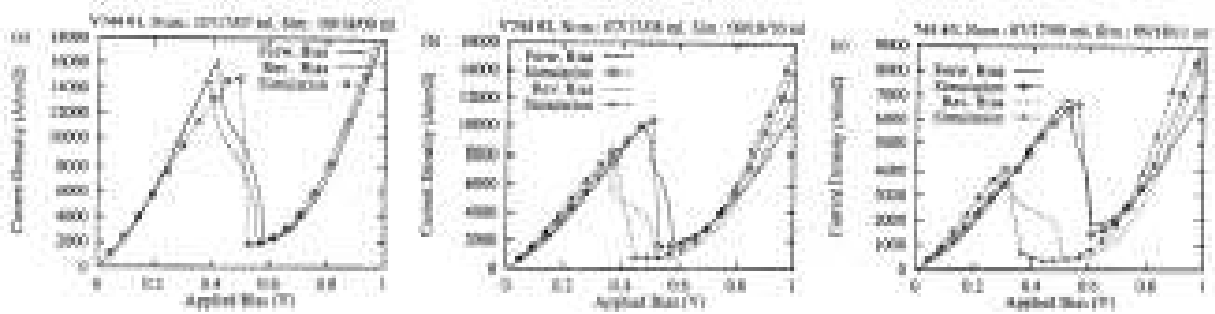} }
\end{center}
%\caption{\mbox{  }}
\caption{Test matrix of In$_{0.47}$Ga$_{0.53}$As / AlAs RTDs.
The collector barrier is increased by one monolayer for the 3 RTDs from left to right. Forward and reverse bias current voltage curves are shown and overlaid on experimental data.
Reprinted from \cite{DRC97_proceedings}. Copyright 1997 IEEE. }
\label{fig:DRC97_fig1a}
\end{figure}

\begin{figure}[h]
\begin{center}
\resizebox{6.5in}{!}{ \includegraphics{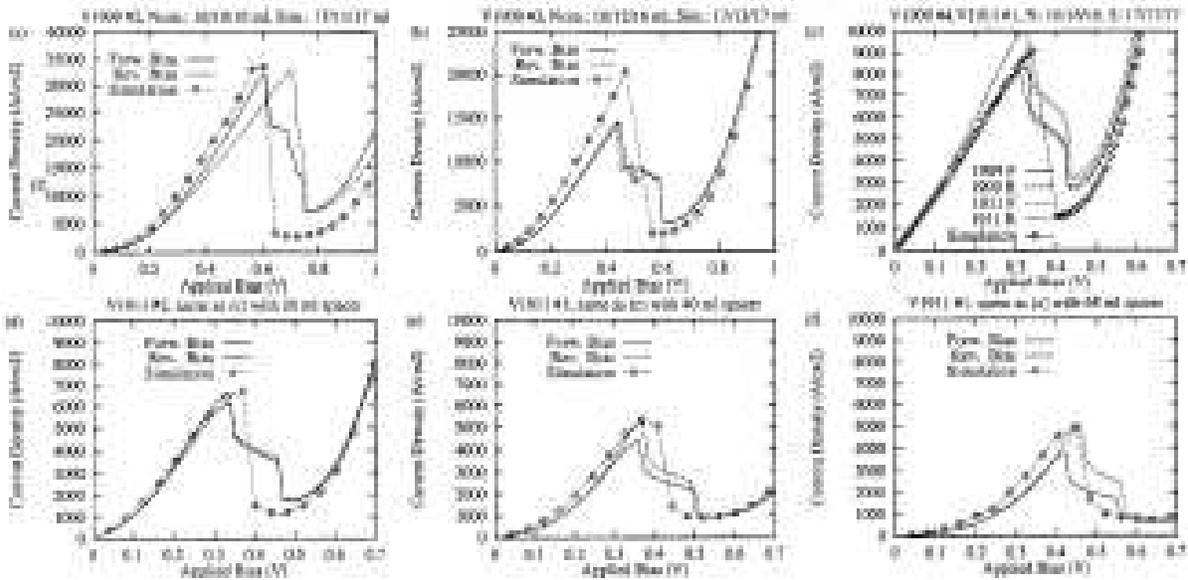} }
\end{center}
%\caption{\mbox{  }}
\caption{Test matrix of In$_{0.47}$Ga$_{0.53}$As / In$_{0.48}$Al$_{0.52}$As RTDs.
(a-c) The well thickness is increased. (c-f) the undoped spacer layer is increased.
Reprinted from \cite{DRC97_proceedings}. Copyright 1997 IEEE.}
\label{fig:DRC97_fig2}
\end{figure}

For room temperature RTDs, simulations indicated that the valley current
was determined by thermionic emission through the second resonance for a number
of RTDs in the test matrix. A comparison of full-band simulations with experimental
data from the test matrix for In$_{0.47}$Ga$_{0.53}$As / AlAs
RTDs and In$_{0.47}$Ga$_{0.53}$As /
In$_{0.48}$Al$_{0.52}$As RTDs is shown in Figs. (\ref{fig:DRC97_fig1a}) and
(\ref{fig:DRC97_fig2}). The simulations model purely coherent transport and yet
match well with the valley current of the experimental data indicating that at
room temperature in these devices, the current is not limited by incoherent
scattering. Purely coherent tunneling calculations were never able to match
the valley current of 'notched well' RTDs with In$_{0.47}$Ga$_{0.53}$As / InAs /
In$_{0.47}$Ga$_{0.53}$As wells \cite{Toms_MIT_RTD}.
These devices, which have the highest peak-to-valley current
ratio of any semiconductor
RTD,
appear to have their room temperature valley current limited by
incoherent scattering.

\begin{figure}[h]
\begin{center}
\resizebox{3.5in}{!}{ \includegraphics{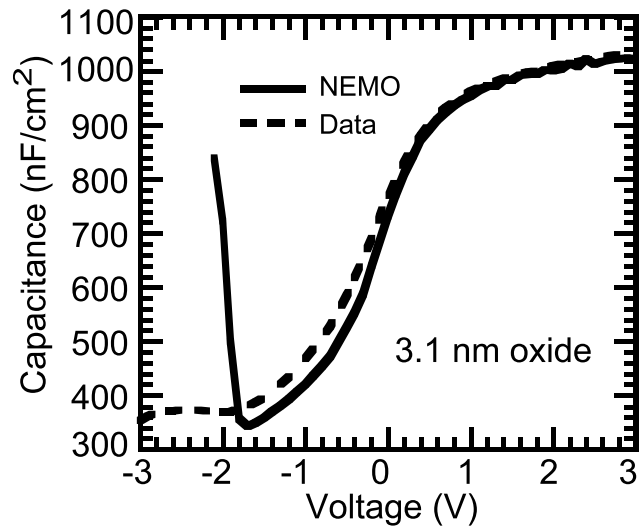} }
\end{center}
%\caption{\mbox{  }}
\caption{Experimental data and simulation of the capacitance - voltage curve of
a MOS structure with a 3.1 nm oxide.  Reprinted from \cite{Lake_IEDM01}. Copyright 2001 IEEE.}
\label{fig:IEDM01_CV}
\end{figure}

As the NEMO program progressed, the simulation software was enhanced to
model other devices and material systems to support the various
experimental programs within TI and Raytheon. One of
the important enhancements was the ability to model the Si / SiGe / SiO$_2$
material system.
Fig. (\ref{fig:IEDM01_CV}) shows close agreement between
the experimental and simulated capacitance-voltage curve of
a metal-oxide-Si (MOS) structure of Brar et al. \cite{Brar96}
The approach taken for Si for quick design calculations as
shown in Fig. (\ref{fig:IEDM01_CV}) was to implement
multiple de-coupled single-band models.
In these models,
Schr\"{o}dinger's equation is solved independently for each band using
Green function techniques.
The quantum charge from each calculation is added, and the total is
used in the Poisson solver.
The quantum calculations and the Poisson calculation are iterated until
convergence using a Newton-Raphson algorithm.
The simulations of the C-V curve used a 4-independent band model consisting of one
band for the 4 equivalent X-valleys, one band for the 2 equivalent X valleys,
one band for the light holes and one band for the heavy holes.

\begin{figure}[h]
\begin{center}
\resizebox{3.5in}{!}{ \includegraphics{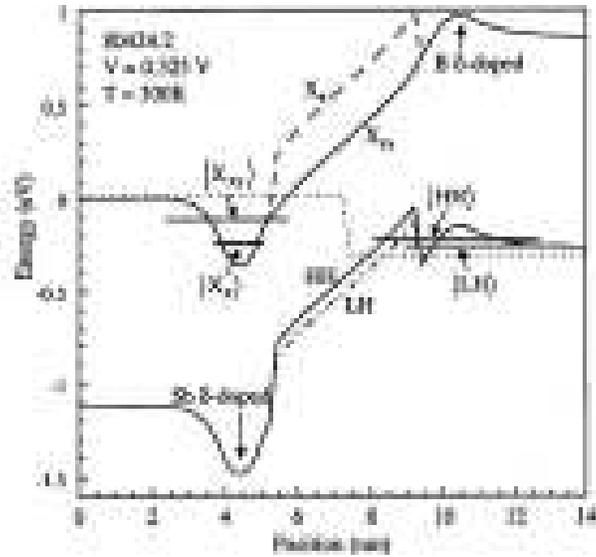} }
\end{center}
%\caption{\mbox{  }}
\caption{Simulation of band profile of the Si / Si$_{0.5}$Ge$_{0.5}$ tunnel diode. Reprinted from \cite{Rommel_APL98}.}
\label{fig:APL1_TD}
\end{figure}

One of the great successes of the NEMO software was its aid in the
design of the first working Si / SiGe tunnel diode \cite{Rommel_APL98}.
There had been significant experimental effort building and
testing devices with designs consisting of an 8 nm Si$_{0.5}$Ge$_{0.5}$
tunnel region
sandwiched between Si.
There was delta doping in the Si on either side of the Si$_{0.5}$Ge$_{0.5}$
tunnel region and the heavy doping continued through the tunnel
region with the n-p junction occuring in the center of the
Si$_{0.5}$Ge$_{0.5}$.
Several iterations with NEMO lead to a new design with the length
of the Si$_{0.5}$Ge$_{0.5}$ tunnel region cut in half from 8 nm to 4nm,
and with the doping completely removed from the
Si$_{0.5}$Ge$_{0.5}$ tunnel region.
The simulations showed that the delta doping on either side
of the 4 nm tunnel region was sufficient to contain
the electric field within the tunnel region and the
band overlap between the n-conduction band and the p-valence band.
The final design simulation is shown in Fig. (\ref{fig:APL1_TD}).
A decoupled multiple single band model was used. In the
strained Si$_{0.5}$Ge$_{0.5}$ region, the multiple bands are
explicitly apparent since they are split in energy by the strain.

\subsubsection{Derivation of the Self-Energies}

Ever since its initial publication,
there have been continued questions concerning the derivation
of the self-energies described in the appendices of
\cite{LakeNemoTheory}. For this reason, we will describe the details
of the derivation of the polar-optical phonon self-energy given
in Appendix A of \cite{LakeNemoTheory}.
%Readers not interested in this level of detail are advised
%to skip to the following section.

We start with the general form of the electron-phonon potential,
\begin{equation}
V_{ep} = \frac{1}{\sqrt{V}}
\sum_{\qv} U_{\qv}
e^{i \qv \cdot \rv}
\left( a_{\qv} + a_{\qv}^{\dagger} \right)
\label{Vep}
\end{equation}
where $\qv$ lies within the first Brillouin zone, $\rv$ is the
electron coordinate, $a$ and $a^{\dagger}$ are the phonon
annihilation and creation operators, and $U_{\qv}$
contains the Fourier transform of the electron-ion potential.
Details of the derivation of this form of $V_{ep}$
can be found in Sec. 1.3 of \cite{MAHANBOOK}.

To second quantize the electron coordinate $\rv$,
we need to define the planar orbital basis and
the electron field operators.
The planar orbital basis for a zincblende (or diamond) lattice
with 2 atoms per unit cell is
\begin{eqnarray}
\ket{a_i, L, \kv} &=& \frac{1}{\sqrt{N}} \sum_{\Rv_t^L} e^{i \kv \cdot \Rv_t^L}
\ket{a_i, L, \Rv_t^L}
\nonumber
\\
\ket{c_i, L, \kv} &=& \frac{1}{\sqrt{N}} \sum_{\Rv_t^L} e^{i \kv \cdot (\Rv_t^L + \vv_t)}
\ket{c_i, L, \Rv_t^L}  .
\label{ack_basis}
\end{eqnarray}
The anions sit on the Bravais lattice and the cations are offset by the
vector $\vv = \frac{a}{4} [111]$. In the planar orbital basis,
$L$ is the layer index in the [001] direction where a layer
includes a layer of anions and a layer of cations
and the layer thickness is $a/2$,
$\Rv_t^L$ is the coordinate in the $x-y$ plane of the
anion in layer $L$, $\Rv_t^L + \vv_t$ is the
coordinate of the corresponding cation in the
$x-y$ plane,
and $\kv$ is a 2 dimensional wavevector in $x$ and $y$.
The indices $a_i$ and $c_i$ label the atomic-like orbitals
s, p$^3$, d$^5$, s* on the anions and cations, respectively.
With this basis, the electron field operators are then defined as
\begin{equation}
\psi(\rv) = \sum_{\kv, L}
\left[
\sum_{a_i} \braket{\rv}{a_i, L, \kv} c_{a_i, L, \kv} + \sum_{c_i} \braket{\rv}{c_i,L,\kv} c_{c_i,L,\kv}
\right]
\label{psi_def}
\end{equation}
where $c_{a_i, L, \kv}$ is the destruction operator for state $\ket{a_i, L, \kv}$
and $c_{c_i,L,\kv}$ is the destruction operator for state $\ket{c_i,L,\kv}$.

The second quantized electron-phonon Hamiltonian is then,
\begin{eqnarray}
\lefteqn{
\hat{H}_{ep} = \frac{1}{\sqrt{V}}
\sum_{\qv} U_{\qv}
\left( a_{\qv} + a_{-\qv}^{\dagger} \right)
\int d^3 r \psi^{\dagger}(\rv) e^{i \qv \cdot \rv} \psi(\rv)
}
\nonumber \\
&&
\;\;\;\;
= \frac{1}{\sqrt{V}} \sum_{\qv} U_{\qv} \left( a_{\qv} + a_{-\qv}^{\dagger} \right)
\sum_{\kv, \kv'} \sum_{L,L'}
\nonumber \\
&&
\;\;\;\;\;\;\;\;\;\;\;\;\;\;\;\;\;\;\;\;
\;\;\;\;\;\;\;\;\;\;\;\;\;\;\;\;\;\;\;\;
\left[
\sum_{a,a'} \bra{a,L,\kv} e^{i\qv \cdot \rv} \ket{a', L', \kv'} c^{\dagger}_{a,L,\kv} c_{a',L',\kv'}
\right.
\label{Hep_i}
\\
&&
\;\;\;\;\;\;\;\;\;\;\;\;\;\;\;\;\;\;\;\;
\;\;\;\;\;\;\;\;\;\;\;\;\;\;\;\;\;\;\;\;
+
\sum_{a,c'} \bra{a,L,\kv} e^{i\qv \cdot \rv} \ket{c', L', \kv'} c^{\dagger}_{a,L,\kv} c_{c',L',\kv'}
\label{Hep_ii}
\\
&&
\;\;\;\;\;\;\;\;\;\;\;\;\;\;\;\;\;\;\;\;
\;\;\;\;\;\;\;\;\;\;\;\;\;\;\;\;\;\;\;\;
+
\sum_{c,a'} \bra{c,L,\kv} e^{i\qv \cdot \rv} \ket{a', L', \kv'} c^{\dagger}_{c,L,\kv} c_{a',L',\kv'}
\label{Hep_iii}
\\
&&
\;\;\;\;\;\;\;\;\;\;\;\;\;\;\;\;\;\;\;\;
\;\;\;\;\;\;\;\;\;\;\;\;\;\;\;\;\;\;\;\;
+
\left.
\sum_{c,c'} \bra{c,L,\kv} e^{i\qv \cdot \rv} \ket{c', L', \kv'} c^{\dagger}_{c,L,\kv} c_{c',L',\kv'}
\right] .
\label{Hep_iv}
\end{eqnarray}
To evaluate the matrix elements of $e^{i \qv \cdot \rv}$ in
(\ref{Hep_i}) - (\ref{Hep_iv}), we must expand out the planar orbitals in terms of the
localized orbitals. The anion-anion matrix elements of (\ref{Hep_i}) become
\begin{equation}
\frac{1}{N} \sum_{\Rv_t, \Rv_t^{\prime}}
\sum_{a,a'}
\sum_{L,L'}
\sum_{\kv, \kv'}
e^{-i\kv \cdot \Rv_t}
\bra{a,L,\Rv_t} e^{i\qv \cdot \rv} \ket{a', L', \Rv_t^{\prime}}
e^{i\kv' \cdot \Rv_t^{\prime}}
c^{\dagger}_{a,L,\kv} c_{a',L',\kv'}
\label{Hep_i_localized_orbital}
\end{equation}
In the long wavelength approximation,
the matrix element in
(\ref{Hep_i_localized_orbital})
is evaluated as
\begin{equation}
\bra{a,L,\Rv_t} e^{i\qv \cdot \rv} \ket{a', L', \Rv_t^{\prime}} =
e^{i q_z L \Delta} e^{i \qv_t \cdot \Rv_t} \delta_{L,L'} \delta_{a,a'} \delta_{\Rv_t, \Rv_t^{\prime}}
\label{Hep_i_matrix_eval}
\end{equation}
where $\Delta = a/2$ is the monolayer (anion plus cation layer)
thickness.
The phase $e^{i \qv \cdot \rv}$ is
evaluated at the position of the anion
$(\Rv_t, L\Delta \hat{\bf z})$
and pulled outside of the integral.
The orthogonality of the orbitals results in the
Kronecker delta functions.
Boykin has shown that, in general, $\rv$ is diagonal in the empirical tight-binding basis, so that
the evaluation of the matrix element in (\ref{Hep_i_matrix_eval}) is true in general and does
not rely on the long wavelength approximation \cite{Boykin_eiqr}.
With the matrix evaluated as in (\ref{Hep_i_matrix_eval}),
term (\ref{Hep_i}) becomes
\begin{eqnarray}
\lefteqn{
\!\!\!\!\!\!\!\!\!\!\!\!\!\!\!\!\!
\frac{1}{N}
\sum_{a}
\sum_{L}
\sum_{\kv, \kv'}
\underbrace{
\sum_{\Rv_t}
e^{\Rv_t \cdot (\qv_t + \kv' - \kv)}
}_{N \delta_{\kv', \kv - \qv_t} }
e^{i q_z L \Delta} c^{\dagger}_{a,L,\kv} c_{a,L,\kv'}
}
\nonumber \\
&&
=
\sum_a \sum_L \sum_{\kv} e^{i q_z L \Delta} c^{\dagger}_{a,L,\kv} c_{a,L,\kv-\qv_t}
\label{Hep_i_final}
\end{eqnarray}
Evaluating the matrix elements as in (\ref{Hep_i_matrix_eval}),
there will clearly be no matrix elements between anion and cation
orbitals, so that terms (\ref{Hep_ii}) and (\ref{Hep_iii}) are zero.
Following the same procedure to evaluate term (\ref{Hep_iv})
gives
\begin{equation}
\sum_c \sum_L \sum_{\kv} e^{i q_z \Delta (L  + 1/2)} c^{\dagger}_{c,L,\kv} c_{c,L,\kv-\qv_t} .
\label{Hep_iv_final}
\end{equation}
The final general form for $\hat{H}_{ep}$ is
\begin{eqnarray}
\lefteqn{
\!\!\!\!\!\!\!\!\!\!\!
\hat{H}_{ep} = \frac{1}{\sqrt{V}}
\sum_{\qv} U_{\qv}
\left( a_{\qv} + a_{-\qv}^{\dagger} \right)
}
\nonumber \\
&&
\;\;\;\;\;\;\;
\sum_L \sum_{\kv}  e^{i q_z L \Delta}
\left[
\sum_a c^{\dagger}_{a,L,\kv} c_{a,L,\kv-\qv_t}
+
e^{i q_z \Delta/2} \sum_c c^{\dagger}_{c,L,\kv} c_{c,L,\kv-\qv_t}
\right] .
\label{Hep_2nd_general}
\end{eqnarray}

To calculate the self-energy, we calculate the path ordered Green
function expanding out the S-matrix in the interaction representation,
\begin{equation}
G^P_{\alp L,\alp'L'}(\kv;t,t')\;=\;
-i\langle {\rm P}\, e^{\frac{-i}{\hbar}\int_c ds H'(s)} c_{\alp,L,\kv}(t) c^\dagger_{\alp',L',\kv}(t') \rangle .
\label{GP}
\end{equation}
P is the path ordering operator, $c$ is the Keldysh contour,
and $H'(s)$ is the perturbing Hamiltonian, in this case, $\hat{H}_{ep}$.
The brackets $\langle \cdots \rangle$ indicate the
nonequilibrium ensemble average \cite{Langreth,Selman_negf}.
Expanding out the exponential to second order, gives two non-zero
terms, the zero order term and the second order term.  The first
order term is zero since
in the absence of interactions,
$\langle a_{\qv} \rangle\,=\,\langle a^{\dagger}_{\qv} \rangle\,=\,0$.

The second order term is
\begin{eqnarray}
\lefteqn{
\left( \frac{-i}{\hbar} \right)^3
\frac{1}{2}
\frac{1}{V}
\sum_{\qv_1,\qv_2}
U_{\qv_1} U_{\qv_2}
\sum_{L_1, L_2}
\sum_{\kv_1, \kv_2}
\sum_{\alp_1,\alp_2}
e^{i q_{z_1} \Delta (L_1 + \xi_{\alp_1})}
e^{i q_{z_2} \Delta (L_2 + \xi_{\alp_2})}
}
\nonumber \\
&&
\;\;\;\;\;\;\;
\langle
P \int_c ds_1 \int_c ds_2
\left(
a_{\qv_1} (s_1) + a_{-\qv_1}^{\dagger}(s_2)
\right)
\left(
a_{\qv_2} (s_2) + a_{-\qv_2}^{\dagger}(s_2)
\right)
\nonumber \\
&&
\;\;\;\;\;\;\;
c^{\dagger}_{\alp_1,L_1,\kv_1}(s_1)
c_{\alp_1,L_1,\kv_1-\qv_{t1}}(s_1)
c^{\dagger}_{\alp_2,L_2,\kv_2}(s_2)
c_{\alp_2,L_2,\kv_2-\qv_{t2}}(s_2)
c_{\alp,L,\kv}(t)
c^{\dagger}_{\alp',L',\kv}(t')
\rangle
\label{GP_2nd_order_term}
\end{eqnarray}
where
$\xi_{\alp}$ is $1/2$ if $\alp$ is a cation orbital and zero otherwise.

We now apply Wick's Theorem to factor the operators and evaluate them in the absence
of interactions \cite{Craig,FetterWalecka}.
The phonon operators become
\begin{eqnarray}
\lefteqn{
\langle P
\left(
a_{\qv_1} (s_1) + a_{-\qv_1}^{\dagger}(s_2)
\right)
\left(
a_{\qv_2} (s_2) + a_{-\qv_2}^{\dagger}(s_2)
\right)
\rangle
}
\nonumber \\
&&
\;\;\;\;\;\;\;
=
\langle
P a_{\qv_1} (s_1) a_{-\qv_2}^{\dagger}(s_2) \rangle
+
\langle P a_{\qv_2} (s_2) a_{-\qv_1}^{\dagger}(s_1) \rangle
\propto
\delta_{\qv_1, -\qv_2} .
\end{eqnarray}
The Kronecker delta function arises because the operators
are evaluated in the absence of interactions so that the
phonon modes are conserved.

There are 2 possible ways to factor the electron operators
to give the rainbow diagram, Fig. (\ref{fig:rainbow}), leading to the standard
self-consistent Born approximation. Both factorizations
are equivalent canceling the factor of $1/2$
in Eq. (\ref{GP_2nd_order_term}).
There is also a bubble diagram identical in form to the Hartree
self-energy diagram. In bulk, due to momentum considerations,
this diagram is exactly zero (p. 401 of Fetter and Walecka \cite{FetterWalecka}).
In a heterostructure, this term
is not necessarily zero, and it has been investigated in detail
by Hyldgaard et al. \cite{Hyldgaard_Ann}.
We have ignored the bubble diagram and only consider the two equivalent
factorizations leading to the self-consistent Born approximation.
Choosing one of the two equivalent factorizations, the electron operators
become
\begin{eqnarray}
\lefteqn{
\!\!\!\!\!\!\!\!
\langle P c_{\alp,L,\kv}(t) c^{\dagger}_{\alp_1,L_1,\kv_1}(s_1) \rangle
\langle P c_{\alp_1,L_1,\kv_1-\qv_{t1}}(s_1) c^{\dagger}_{\alp_2,L_2,\kv_2}(s_2) \rangle
\langle P c_{\alp_2,L_2,\kv_2-\qv_{t2}}(s_2) c^{\dagger}_{\alp',L',\kv}(t') \rangle
}
\nonumber \\
&&
\!\!\!\!
=
\delta_{\kv,\kv_1} g^P_{\alp L,\alp_1 L_1}(\kv;t,s_1)
\delta_{\kv_2,\kv_1-\qv_{t1}} g^P_{\alp_1 L_1,\alp_2 L_2}(\kv_1 - \qv_{t1};s_1,s_2)
\delta_{\kv_2-\qv_{t2},\kv} g^P_{\alp_2 L_2,\alp' L'}(\kv;s_2,t')
\label{electron_ops_factored}
\end{eqnarray}
In (\ref{electron_ops_factored}), a lower case $g$ is used to indicate the
bare Green function in the absence of the electron-phonon interaction.
Since the operators are evaluated in the absence of phonon interactions,
transverse momentum is conserved giving rise to the Kronecker delta
functions.

With the factorization of the electron and phonon operators, the second order term
(\ref{GP_2nd_order_term}) becomes
\begin{eqnarray}
\lefteqn{
\!\!\!\!\!\!\!\!
\!\!\!\!\!\!\!\!
\!\!\!\!\!\!\!\!
\frac{1}{V}
\sum_{\qv}
\left| U_{\qv} \right|^2
\sum_{L_1, L_2}
\sum_{\alp_1,\alp_2}
e^{i q_{z} \Delta (L_1 - L_2 + \nu_{\alp_1,\alp_2})}
}
\nonumber \\
&&
\int_c \! ds_1 \! \int_c \! ds_2
\left[
\langle P a_{\qv} (s_1) a_{\qv}^{\dagger}(s_2) \rangle
+
\langle P a_{-\qv} (s_2) a_{-\qv}^{\dagger}(s_1) \rangle
\right]
\nonumber \\
&&
\;\;\;\;\;\;\;\;\;\;\;\;\;\;\;\;\;\;
g^P_{\alp L,\alp_1 L_1}(\kv;t,s_1)
g^P_{\alp_1 L_1,\alp_2 L_2}(\kv - \qv_t;s_1,s_2)
g^P_{\alp_2 L_2,\alp' L'}(\kv;s_2,t')
\label{GP_2nd_order}
\end{eqnarray}
where
\begin{equation}
\nu_{\alpha_1, \alpha_2} \;=\;\left\{ \begin{array}{ll}
\frac{1}{2} & \mbox{$\alpha_1=c$, $\alpha_2 = a$} \\
-\frac{1}{2} & \mbox{$\alpha_1=a$, $\alpha_2 = c$} \\
0 & \mbox{otherwise}
\end{array}
\right.
\label{PRB1_nu}
\end{equation}

The general form of the path ordered Dyson's equation resulting from expanding
Eq. (\ref{GP}) is
\begin{eqnarray}
\lefteqn{
G^P_{\alp L,\alp'L'}(\kv;t,t')\;=\;
g^P_{\alp L,\alp'L'}(\kv;t,t')
}
\nonumber \\
&&
+\:\int_c ds \int_c ds'
\sum_{L_1,L_2}
\sum_{\alp_1,\alp_2}
g^P_{\alp L, \alp_1 L_1}(\kv;t,s) \Sigma^P_{\alp_1 L_1, \alp_2 L_2}(\kv;s,s')
g^P_{\alp_2 L_2, \alp' L'}(\kv;s',t')
\label{P_dyson}
\end{eqnarray}
Comparing Eq. (\ref{P_dyson}) to Eq. (\ref{GP_2nd_order}), we identify the self-energy as
\begin{eqnarray}
\lefteqn{
\!\!\!\!\!\!\!\!\!\!\!\!\!\!\!\!\!\!\!\!\!
\Sigma^P_{\alp_1 L_1, \alp_2 L_2} (\kv; s_1,s_2)
=
\frac{1}{V}
\sum_{\qv}
\left| U_{\qv} \right|^2
e^{i q_{z} \Delta (L_1 - L_2 + \nu_{\alp_1,\alp_2})}
}
\nonumber \\
&&
\left[
\langle P a_{\qv} (s_1) a_{\qv}^{\dagger}(s_2) \rangle
+
\langle P a_{-\qv} (s_2) a_{-\qv}^{\dagger}(s_1) \rangle
\right]
G^P_{\alp_1 L_1,\alp_2 L_2}(\kv - \qv_t;s_1,s_2)
\label{SigmaP_general}
\end{eqnarray}
In going from Eq. (\ref{GP_2nd_order}) to Eq. (\ref{SigmaP_general}),
we have replaced the bare $g^P$ with the full $G^P$ that must
be calculated self-consistently with the self-energy $\Sigma^P$.
This replacement of $g^P$ with $G^P$ sums the whole class of
rainbow diagrams an example of which is shown in the context of
Hartree-Fock theory in Fig. 10.2 of Fetter and Walecka \cite{FetterWalecka}.

The real-time self energies that we need, $\Sigma^<$, $\Sigma^>$,
and $\Sigma^R$, are obtained from $\Sigma^P$
by considering the individual cases of $s_1$ and $s_2$
lying on different branches of the Keldysh contour.
Or, one can simply use the relations of Langreth \cite{Langreth}.
For pedagogical reasons, we will evaluate $\Sigma^<$
by considering the time contour and not using the Langreth rules.
$\Sigma^<(s_1,s_2)$ is obtained from $\Sigma^P(s_1,s_2)$
by considering the case of $s_2 > s_1$. According to the
path ordering operator $P$, the only way that $s_2$
can always be greater than $s_1$ is if $s_2$ lies on the
upper branch of the contour and $s_1$ lies on the
lower branch.
Applying the path ordering operator, which places operators of later times to the left,
to the time dependent terms of Eq. (\ref{SigmaP_general}), we obtain
\begin{equation}
\left[
\langle  a_{\qv}^{\dagger}(s_2) a_{\qv}(s_1) \rangle
+
\langle a_{-\qv} (s_2) a_{-\qv}^{\dagger}(s_1) \rangle
\right]
G^<_{\alp_1 L_1,\alp_2 L_2}(\kv - \qv_t;s_1,s_2)
\label{eq:less_than_ordering}
\end{equation}
where
\begin{equation}
G^<_{\alp_1 L_1,\alp_2 L_2}(\kv - \qv_t;s_1,s_2) =
\frac{i}{\hbar} \langle
c^{\dagger}_{\alp_2,L_2,\kv -\qv_t}(s_2)
c_{\alp_1,L_1,\kv-\qv_{t}}(s_1)
\rangle .
\label{eq:G<_def}
\end{equation}
The operators $a$ and $a^{\dagger}$ are evaluated in the absence
of interactions. Therefore,
their time dependence is
\begin{eqnarray}
a_{\qv}(s)&=&a_{\qv} e^{-i \w_{\qv} s}
\nonumber \\
a^{\dagger}_{\qv}(s)&=&a^{\dagger}_{\qv} e^{i \w_{\qv} s} .
\label{eq:a_time_dependence}
\end{eqnarray}
The phonon ensemble averages are evaluated as
\begin{eqnarray}
\langle  a_{\qv}^{\dagger} a_{\qv} \rangle&=&n_{\qv}
\nonumber \\
\langle a_{\qv} a_{\qv}^{\dagger} \rangle&=&n_{\qv} + 1
\label{eq:phonon_number_eval}
\end{eqnarray}
where $n_{\qv}$ is the average phonon number in mode $\qv$ which,
in equilibrium, is the Bose-Einstein factor.
Using (\ref{eq:phonon_number_eval}) and (\ref{eq:a_time_dependence}),
the time dependent terms of (\ref{eq:G<_def}) become
\begin{equation}
\left[
n_{\qv} e^{-i \w_{\qv} (s_1 - s_2)} + \left( n_{\qv} + 1 \right)
e^{i \w_{\qv} (s_1 - s_2)}
\right]
G^<_{\alp_1 L_1,\alp_2 L_2}(\kv - \qv_t;s_1,s_2)
\label{eq:t_dependence_less_than}
\end{equation}
In steady state, $G^<$ is a function of the time difference coordinate
$s_1 - s_2$ which we Fourier transform into energy
according to
\begin{eqnarray}
\lefteqn{
\!\!\!\!\!\!\!\!\!\!\!\!\!\!\!
\int d(s_1 - s_2)
e^{i \frac{E}{\hbar} (s_1 - s_2)}
\left[
n_{\qv} e^{-i \w_{\qv} (s_1 - s_2)} + \left( n_{\qv} + 1 \right)
e^{i \w_{\qv} (s_1 - s_2)}
\right]
G^<_{\alp_1 L_1,\alp_2 L_2}(\kv - \qv_t;s_1-s_2)
}
\nonumber \\
&&
=
n_{\qv} G^<_{\alp_1 L_1,\alp_2 L_2}(\kv - \qv_t;E-\hbar \w_{\qv})
+
\left( n_{\qv} + 1 \right) G^<_{\alp_1 L_1,\alp_2 L_2}(\kv - \qv_t;E+\hbar \w_{\qv}) .
\label{eq:E_dependence_less_than}
\end{eqnarray}
Placing this back into Eq. (\ref{SigmaP_general}), we obtain the general form for the
phonon scattering self-energy within the self-consistent Born
approximation.
\begin{eqnarray}
\lefteqn{
\!\!\!\!\!\!\!\!\!\!\!\!\!
\Sigma^<_{\alp_1 L_1, \alp_2 L_2} (\kv; E)
=
\frac{1}{V}
\sum_{\qv}
\left| U_{\qv} \right|^2
e^{i q_{z} \Delta (L_1 - L_2 + \nu_{\alp_1,\alp_2})}
}
\nonumber \\
&&
\left[
n_{\qv} G^<_{\alp_1 L_1,\alp_2 L_2}(\kv - \qv_t;E-\hbar \w_{\qv})
+
\left( n_{\qv} + 1 \right) G^<_{\alp_1 L_1,\alp_2 L_2}(\kv - \qv_t;E+\hbar \w_{\qv})
\right]
\label{Sigma<_general}
\end{eqnarray}

The evaluation of $\Sigma^>$ proceeds in exactly the same way
with the difference being that $s_1 > s_2$ when evaluating
the time-dependent terms of
Eq. (\ref{SigmaP_general}).
Therefore, the operator ordering is reversed
in Eq. (\ref{eq:less_than_ordering}).
The result is
\begin{eqnarray}
\lefteqn{
\!\!\!\!\!\!\!\!\!\!\!\!\!
\Sigma^>_{\alp_1 L_1, \alp_2 L_2} (\kv; E)
=
\frac{1}{V}
\sum_{\qv}
\left| U_{\qv} \right|^2
e^{i q_{z} \Delta (L_1 - L_2 + \nu_{\alp_1,\alp_2})}
}
\nonumber \\
&&
\left[
n_{\qv} G^>_{\alp_1 L_1,\alp_2 L_2}(\kv - \qv_t;E+\hbar \w_{\qv})
+
\left( n_{\qv} + 1 \right) G^>_{\alp_1 L_1,\alp_2 L_2}(\kv - \qv_t;E-\hbar \w_{\qv})
\right]
\label{Sigma>_general}
\end{eqnarray}

The relationship of the retarded self-energy $\Sigma^R$ to $\Sigma^<$
and $\Sigma^>$ is
\begin{equation}
\Sigma^R(t,t') = \Theta(t-t')
\left[ \Sigma^>(t,t') - \Sigma^<(t,t') \right] = -i \Theta(t-t')
\Gamma(t,t')
\end{equation}
or, Fourier transforming the time difference coordinate to energy,
\begin{equation}
\Sigma^R(E)\;=\;
{\rm P} \int \frac{dE'}{2\pi}
\frac{\Gamma(E')}{E-E'}
\;-\;\frac{i}{2} \Gamma(E) .
\label{PRB1_SR_Gamma}
\end{equation}
Since $\Gamma(E)$ is Hermitian, the principal value integral is
also Hermitian. This term gives rise to an energy renormalization
but no relaxation or dephasing. Because the principal value
integral is so difficult to perform numerically, this term
is often ignored and $\Sigma^R(E)$ is approximated
as $\frac{-i}{2} \Gamma(E)$ where
\begin{equation}
\Gamma(E)\;=\;
i \left[ \Sigma^>(E) - \Sigma^<(E) \right].
\label{PRB1_Gamma_def}
\end{equation}

Another approach, the one implemented in NEMO and described in
\cite{LakeNemoTheory},
is to use the single-electron or low-density approximation.
In this approximation
$\Sigma^<$ is set to zero since the inscattering into
energy $E$ is proportional to the electron density $G^<$
at energies $E \pm \hbar \w$. The
electron density is assumed low and thus approximated as
equal to zero.
The outscattering $\Sigma^>$ at energy E is proportional to the
hole density $G^>$ at energies $E \pm \hbar \w$. Since the
electron density is zero, the hole density is just the total
density of states. Therefore $G^>$ is
replaced in Eq. (\ref{Sigma>_general}) by $-iA$ where
$A$ is the spectral function.
All of the energy dependence of $\Sigma^>$ is contained in the
spectral function. Furthermore, from Eq. (\ref{PRB1_Gamma_def}),
$\Sigma^> = -i\Gamma$.
Since $G^R$ is related to $A$ in exactly the
same way as $\Sigma^R$ is related to $\Gamma$,
Eq. (\ref{PRB1_SR_Gamma}) converts the spectral functions
occurring in $\Sigma^>$ into retarded Green functions.
The retarded self energy then has exactly the same form
as $\Sigma^>$ Eq. (\ref{Sigma>_general}) with $G^R$ replacing $G^>$.
This form of the self-energy does conserve current.

With the general form of the self-energy (\ref{Sigma>_general}),
we now consider the specific form of $U_{\qv}$ corresponding to
dispersionless polar optical phonons.
For dispersionless polar optical phonons, $\w_{\qv} = \w_o$,
$n_{\qv} = n_B(\hbar \w_o) = 1/(e^{\hbar \w_o/k_B T} - 1)$,
and
\begin{equation}
|U_{\bf q}|^2\:=\:
\frac{e^2 \hbar\omega_o}{2}
\left( \frac{1}{\epsilon_\infty} -\frac{1}{\epsilon_o} \right)
\frac{q^2}{(q^2 + q_o^2)^2}
\end{equation}
where $q_o$ is the inverse screening length \cite{Ridley_book}.
For numerical reasons, we prefer to have $G^>$ in Eq. (\ref{Sigma>_general})
be only a function of $\qv_t$ rather than
$\kv - \qv_t$. Therefore, we change dummy indices so that
the expression for $\Sigma^>$ becomes
\begin{eqnarray}
\lefteqn{
\!\!\!\!\!\!\!\!\!\!\!\!\!
\Sigma^>_{\alp_1 L_1, \alp_2 L_2} (\kv; E)
=
\frac{1}{V}
\sum_{\qv}
\left| U_{\kv - \qv} \right|^2
e^{-i q_{z} \Delta (L_1 - L_2 + \nu_{\alp_1,\alp_2})}
}
\nonumber \\
&&
\left[
n_B G^>_{\alp_1 L_1,\alp_2 L_2}(\qv_t;E+\hbar \w_{o})
+
\left( n_{B} + 1 \right) G^>_{\alp_1 L_1,\alp_2 L_2}(\qv_t;E-\hbar \w_{o})
\right]
\label{Sigma>_pop}
\end{eqnarray}

When it is appropriate to approximate $G^>(\qv_t;E)$ as being independent
of the angle of $\qv_t$ in the $x-y$ plane, then we can perform
the angular integration analytically. This approximation would be
appropriate near the conduction band edge of GaAs where the bands
are spherical.
The sum over $\qv$ becomes the following integral in
cylindrical coordinates,
\begin{equation}
\int
\frac{dq_t q_t}{(2\pi)^2}
G^>(q_t ;E \pm \hbar \w_{o})
\int_{\frac{-\pi}{\Delta}}^{\frac{\pi}{\Delta}}
\frac{dq_z}{2\pi}
e^{-i q_{z} \Delta (L_1 - L_2 + \nu_{\alp_1,\alp_2})}
\underbrace{
\int_0^{2\pi} \!\!d \theta
\frac{|\kv - \qv |^2}{ \left(|\kv - \qv |^2 + q_o^2 \right)^2 }
}_{I_{\theta}}
\label{eq:d3q_integral}
\end{equation}
The integral over $\theta$ is
\begin{equation}
I_{\theta} =
\int_0^{2\pi} \!\!d \theta
\frac{k^2 + q_t^2 + q_z^2 - 2kq_t \cos \theta}
{ \left( k^2 + q_t^2 + q_z^2 + q_o^2 - 2kq_t\cos \theta \right)^2 } \;\;\;\;.
\label{eq:I_theta}
\end{equation}
This integral is evaluated by making the substitution
\begin{equation}
z = e^{i\theta}
\label{z_sub}
\end{equation}
which converts the integral over $\theta$
into a contour integral around the unit circle.
The integral is then evaluated using the residue theorem.
With substitution (\ref{z_sub}), Eq. (\ref{eq:I_theta}) becomes
\begin{equation}
I_{\theta} =
\frac{i}{kq_t}
\oint dz
\left\{
\frac{1}{z^2 -bz +1} +
\frac{ \frac{q_o^2}{kq_t} z}{ \left[ z^2 -bz +1 \right]^2 }
\right\}
\label{I_z}
\end{equation}
where
\begin{equation}
b = \frac{k^2 + q_t^2 + q_z^2 + q_o^2}{kq_t} .
\label{eq:b}
\end{equation}
Evaluating (\ref{I_z}) with the residue theorem, we obtain
\begin{equation}
I_{\theta}(q_z^2, q_t^2, k^2) =
2\pi
\left\{
\frac{1}{\sqrt{(q_z^2 + q_t^2 + k^2 + q_o^2)^2 \,-\,4k^2 q_t^2} }
-
\frac{q_o^2(q_z^2 + q_t^2 + k^2 + q_o^2)}
{\left[(q_z^2 + q_t^2 + k^2 + q_o^2)^2 \,-\,4k^2 q_t^2\right]^{3/2}}
\right\}
\end{equation}
We write the integral over $q_z$ in Eq. (\ref{eq:d3q_integral}) as
\begin{equation}
2 \int_0^{\pi/\Delta} \frac{dq_z}{2 \pi}
\cos\left[
q_{z} \Delta (L_1 - L_2 + \nu_{\alp_1,\alp_2}
\right]
I_{\theta}(q_z^2, q_t^2, k^2)
\label{eq:dqz}
\end{equation}
and the final form of the self energy becomes
\begin{eqnarray}
\Sigma^<_{\alpha,L; \alpha^{'},L^{'}}(k;E) &=&
\frac{e^2 \hbar\omega_o}{2\pi}
\left( \frac{1}{\epsilon_\infty} -\frac{1}{\epsilon_o} \right)
\int \frac{d^2 q_t}{4\pi^2}
I \left(\left| L-L^{'}+\nu_{\alpha, \alpha^{'}} \right|, k^2, q_t^2 \right) \nonumber
\\
&&\;\;\;\;\;\;\;\;
\left[
n_B G^>_{\alpha,L; \alpha^{'},L^{'}} (q_t; E + \hbar \omega) \;+\;
(n_B+1) G^>_{\alpha,L; \alpha^{'},L^{'}} (q_t; E - \hbar \omega)
\right]
\label{PRB1_pop_s<k}
\end{eqnarray}
where
\begin{eqnarray}
\lefteqn{
I \left(\left|L-L^{'}+\nu_{\alpha, \alpha^{'}} \right|, k^2, q_t^2 \right) =}
\nonumber \\
&&
\;\;\;\;\;\;\;\;\;\;\;
\;\;\;\;\;\;\;\;\;\;\;
\int_0^{\pi/\Delta} dq_z
\frac{\cos\left[q_z \Delta (L-L^{'}+\nu_{\alpha, \alpha^{'}}) \right]}
{\sqrt{(q_z^2 + q_t^2 + k^2 + q_o^2)^2 \,-\,4k^2 q_t^2}}
\nonumber \\
&&
\;\;\;\;\;\;\;\;\;\;\;
\;\;\;\;\;\;\;\;\;\;\;
-
\int_0^{\pi/\Delta} dq_z \frac{\cos\left[q_z \Delta (L-L^{'}+\nu_{\alpha, \alpha^{'}}) \right]
q_o^2(q_z^2 + q_t^2 + k^2 + q_o^2)}
{\left[(q_z^2 + q_t^2 + k^2 + q_o^2)^2 \,-\,4k^2 q_t^2\right]^{3/2}}
\end{eqnarray}

Note that the integrand in Eq. (\ref{PRB1_pop_s<k})
is only a function of the magnitude $q_t = |\qv_t|$.
The integral over the angle was done analytically.
We keep the form of the
full two dimensional integral
because, in an effective mass model,
we make the following change of
variables,
\begin{equation}
\int \frac{d^2 q_t}{4\pi^2}\;\longrightarrow\;
\int_0^B d\varepsilon_{q_t} \rho_{2D}(\varepsilon_{q_t})
\; \longrightarrow \;
\int_{E_{min}}^E dE_z \rho_{2D}(E-E_z)
\label{change_vars}
\end{equation}
where $\varepsilon_{q_t}$ and $E_z$ are defined variables,
$E$ is the total energy, $B$ is the bandwidth,
$\rho_{2D}$ is the 2D density of
states, and $E_{min}$ is a minimum energy
used in the numerical integration.
$\varepsilon_{q_t}$
is the transverse energy using the
dispersion relation in the flat band part of the injecting contact.
For a parabolic
dispersion,
$\varepsilon_{q_t} = \frac{\hbar^2 q_t^2}{2m^{*}}$ where
$m^{*}_L$ is the effective mass in the
injecting contact.
Therefore, $\rho_{2D} = m^{*}/(2 \pi \hbar^2)$.
The longitudinal energy is defined as $E_z = E - \varepsilon_{q_t}$.

\subsection{Post NEMO 1D NEGF Developments}
Since its delivery to the U.S. government in 1997,
the public availability of NEMO has been sporadic \cite{Lake_IEDM01}.
Currently,
NEMO is semi-publicly available with restrictions \cite{gekco_NEMO1D_url}.
Interesting post-NEMO developments of applications
for NEGF to 1D transport through planar semiconductor structures are
the formulation of NEGF theory to model superlattices and quantum cascade
lasers \cite{Lee_Wacker02} and the application of NEGF theory to model
the indirect phonon assisted tunneling in Si
tunnel diodes \cite{CR_APL1,CR_JAP1}.

\subsubsection{Interband, Intervalley Tunneling in Si}
The NEGF theory developed to model
indirect phonon assisted intervalley, interband
tunneling in Si tunnel diodes went back to the
original theory of Caroli et al.
described in the 4th article of the
series \cite{Caroli_IV}.
The original theory was rederived
for localized orbital full-band models
with the specific types of phonons important
for interband tunneling in Si.
A detailed description of the theory is
given in \cite{CR_JAP1}.
Calculations were performed both with a
second nearest neighbor sp$^3$s$^*$ \cite{Boykin96_1}
and a nearest neighbor sp$^3$s$^*$d$^5$
planar orbital basis \cite{Jancu}.
The specific phonons of interest were the
zone edge TA and TO phonons.
From a theoretical and computational point of view,
modeling interband tunneling in Si is
interesting and challenging because it is
primarily an indirect, phonon assisted process.
Furthermore, it is an intervalley process in which
an electron starts out in one of the 6 valleys
near X along the $\Delta$ line and tunnels to
the valence band at $\Gamma$.

\begin{figure}[h]
\begin{center}
\resizebox{3.5in}{!}{ \includegraphics{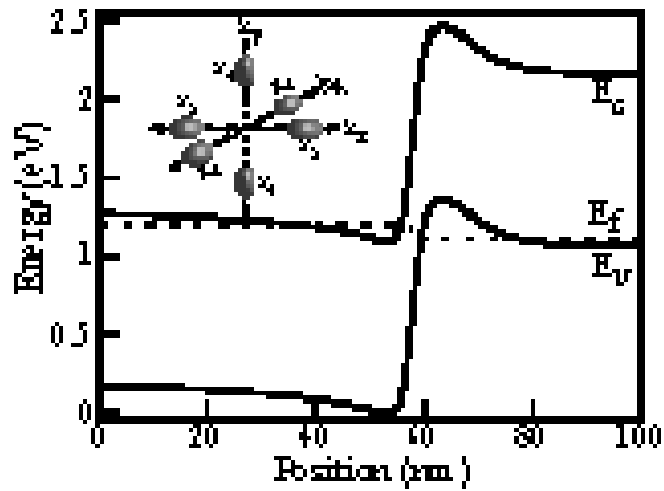} }
\end{center}
%\caption{\mbox{  }}
\caption{ Band profile of the delta-doped Si tunnel diode biased at 0.1V. Inset, 6 X valleys of the conduction band. Reprinted from \cite{CR_APL1}.}
\label{fig:Si_TD_bands_6valleys}
\end{figure}

The band diagram of the device simulated along
with the 6 valleys of the conduction band is
shown in Fig. \ref{fig:Si_TD_bands_6valleys}.
The tunnel current is dominated by the
electrons in the 4 transverse valleys whose
long axis is perpendicular to the direction of current
flow.
In the simplest picture,
this is because the effective mass in the tunneling direction of the 4
transverse valleys is 5 times lighter than that of the 2 longitudinal valleys,
and their combined 2D density of states is 5 times greater.
There is a small coherent tunnel current that arises from electrons
tunneling from the 2 longitudinal valleys of the conduction band
directly to the valence band. This is possible since the transverse
wavevector of the electrons in the 2 longitudinal valleys
is centered at $k=0$ and can therefore be conserved in the
coherent tunneling process.

\begin{figure}[h]
\begin{center}
\resizebox{3.5in}{!}{ \includegraphics{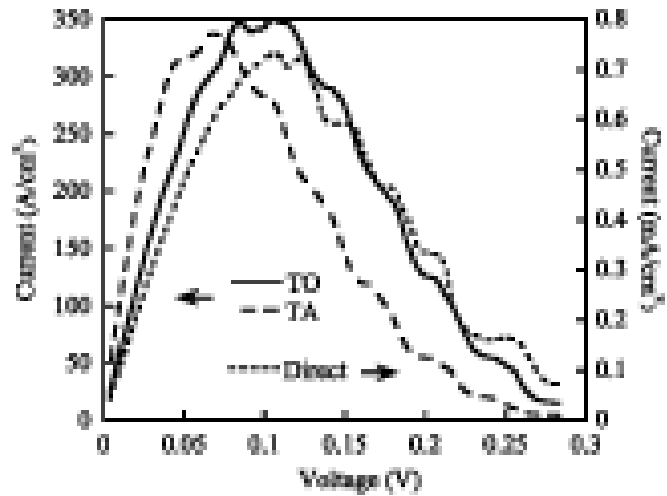} }
\end{center}
%\caption{\mbox{  }}
\caption{Three current components: TA phonon assisted, TO phonon assisted, and
direct or coherent current. Inset, 2D electron and hole dispersions. Reprinted from \cite{CR_APL1}.}
\label{fig:CR_APL1_fig2}
\end{figure}

The current modeled has 3 separate components.
The coherent component is given by the usual expression
\cite{CR_JAP1,FisherLee,DattaTextbook2,LakeNemoTheory}.
\begin{equation}
J = \frac{e}{A \hbar} \sum_{\kv} \int \frac{dE}{2 \pi}
{\rm tr} \left\{ \GL g^R \GR g^A \right\} \left( \fL - \fR \right)
\label{eq:coherentJ}
\end{equation}
In Eq. (\ref{eq:coherentJ}), lower case $g$ symbols are used
for the Green functions since they are the bare Green functions
calculated in the absence of phonon scattering.
The phonon assisted component of the current is given by
\cite{CR_JAP1}
\begin{eqnarray}
J_1 &=&
\frac{4e (D_tK)^2}{\rho \omega a}
\int_{X} \frac{d^2\kvX}{4 \pi^2}
\int_{\G} \frac{d^2\kvG}{4 \pi^2}
\int \frac{dE}{2 \pi}
\sum_n
\left[
{\rm tr}
\left\{
a^{X}_n(\kvX,E)
\right\}
\right.
\nonumber
\\
&&
\left[
{\rm tr}
\left\{
a^{\G}_n(\kvG,E-\hw )
\right\}
\fX(E)
\left( 1-\fG(E-\hw) \right)
(n_B(\hw) + 1)
\right.
\nonumber
\\
&+&
{\rm tr}
\left\{
a^{\G}_n(\kvG,E + \hw ) n_B(\hw)
\right\}
\fX(E) \left(1-\fG(E+\hw ) \right)
\nonumber
\\
&-&
{\rm tr}
\left\{
a^{\G}_n(\kvG,E-\hw )
\right\}
\left( 1 - \fX(E) \right) \fG(E-\hw)
n_B(\hw)
\nonumber
\\
&-&
\left.
\left.
{\rm tr}
\left\{
a^{\G}_n(\kvG,E+\hw )
\right\}
\left( 1 - \fX(E) \right) \fG(E+\hw)
(n_B(\hw)+1)
\right]
\right] .
\label{eq:Caroli49_envelopes}
\end{eqnarray}
Eq. (\ref{eq:Caroli49_envelopes}) is Fermi's Golden
Rule written in Green function form. It is a full-band
version of Eq. (49) of Caroli et al. \cite{Caroli_IV} written for
TO and TA zone edge phonons.
To emphasize the intervalley process, subscripts and superscripts
$\Gamma$ and $X$ are used in Eq. (\ref{eq:Caroli49_envelopes}).
The quantity $a^X$ indicates the left injected spectral
function corresponding to Fig. \ref{fig:Si_TD_bands_6valleys}.
Since it is injected from the left, it is in one of the $X$
valleys of the conduction band.
Similarly, $a^{\Gamma}$ indicates the right injected
spectral function which is in the valence band near $\Gamma$.
The integrations over the transverse momenta are performed
around the respective valley minimum points.
Eq. (\ref{eq:Caroli49_envelopes}) is evaluated individually
for each type of phonon, TO and TA, and the results are added.
The relative magnitude of the 3 components can be seen
in Fig. \ref{fig:CR_APL1_fig2}.
Note that the axis for the ``Direct'' current which is the
coherent current is in mA whereas the axis for the phonon
assisted current is in A. The magnitude of the coherent
current is approximately 5 orders of magnitude smaller than the
phonon assisted current.

One issue that was theoretically investigated
was to understand the effect of quantized
states in the contacts. These formed in the delta-doped
wells shown in Fig. \ref{fig:Si_TD_bands_6valleys}.
The effect of these states was to give some structure to
the current-voltage curves primarily in the
negative differential resistance (NDR) region as is
shown in Fig. \ref{fig:CR_APL1_fig2}.
The inset of  Fig. \ref{fig:CR_APL1_fig2} gives an intuitive
explanation of why the structure occurs in the NDR region.
The 2D channels are initially on. As they uncross, the channels
get turned off giving structure to the I-V.
Since the quantum wells are degenerately doped,
inclusion of realistic broadening in the calculations
completely wipes out any structure in the I-V \cite{CR_APL1}.

\begin{figure}[h]
\begin{center}
\resizebox{3.5in}{!}{ \includegraphics{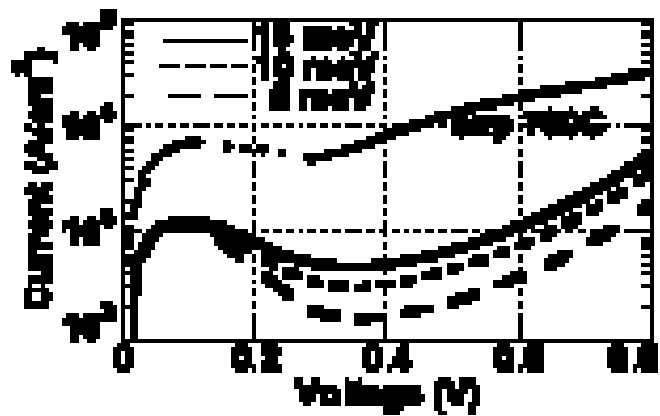} }
\end{center}
%\caption{\mbox{  }}
\caption{Experimental I-V from ref. \cite{Thompson_APL99} corrected for
12 $\Omega$ series resistance overlaid on calculated
I-Vs with different energy broadening in the contacts as
shown in the legend. Reprinted from \cite{CR_JAP1}.}
\label{fig:CR_JAP1_fig6}
\end{figure}

A comparison between the experimental and calculated I-V
curves is shown in Fig. \ref{fig:CR_JAP1_fig6}.
The discrepancy in the magnitude of the experimental and calculated
current (approximately a factor of 5) can be the result
of several factors. The magnitude of the current depends
quadratically on the magnitude of the deformation potential
and exponentially on the width of the tunnel junction and the
evanescant wave vector in the tunnel junction.
A series of calculations were performed increasing the
separation of the delta-doping planes and thus increasing the
tunnel junction width. The magnitude of the peak current
had a dependence of $\exp (-d/0.42)$ where $d$ is the tunnel
junction width in nm.
For reference, the simplest analytic, parabolic, effective mass model
results in a dependence of $\exp (-d/0.5)$.
The dependence of $\exp (-d/0.42)$ agreed
fairly well with the dependence
of $\exp (-d/0.38)$
extracted from reverse biased Si pn junctions from a later
exhaustive experimental study
\cite{PSolomon_JAP_5_04}.

\begin{figure}[h]
\begin{center}
\resizebox{3.5in}{!}{ \includegraphics{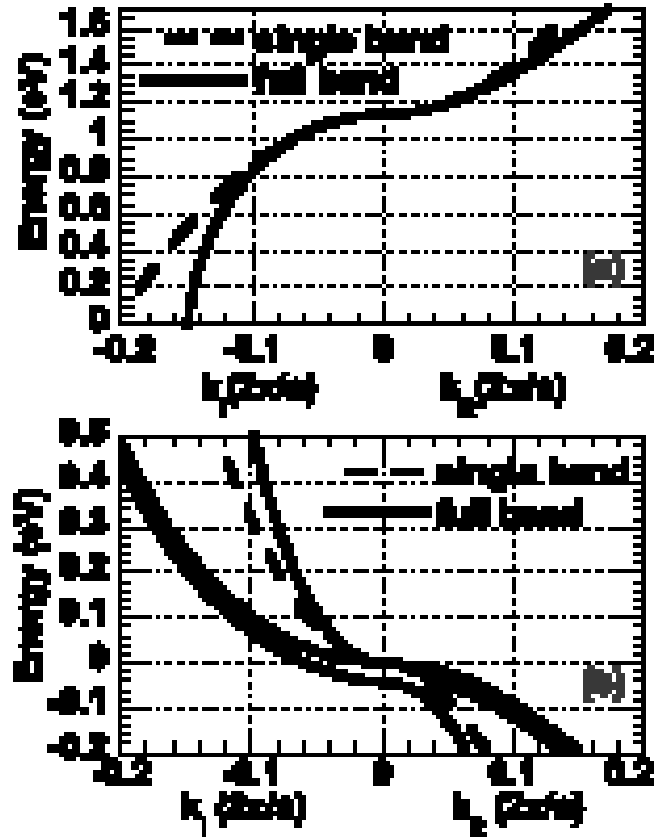} }
\end{center}
%\caption{\mbox{  }}
\caption{Comparison of the real and imaginary dispersion relations calculated
(a) from the parabolic single band model using the transverse conduction band mass of
0.19 m$_0$ and from the full-band model in the
transverse mass direction of the $X_4$ conduction band valley
and (b) from the single band model using the light hole mass of 0.16 m$_0$ and from
the full-band model in the valence band in the (001) direction.
The horizontal axis to the left of $0$ is imaginary $k$ and to the right of $0$ is
real $k$. Reprinted from \cite{CR_JAP1}.}
\label{fig:CR_JAP1_fig7}
\end{figure}

The decay lengths are determined by the evanescent
wavevectors in the bandgap.
The full band calculations are compared with
the parabolic effective mass model in Fig. \ref{fig:CR_JAP1_fig7}.
The full band calculations result in evanescent
wavevectors in the gap that are smaller than that
of the parabolic effective mass model.
This is to be expected since the evanescent hole bands
wrap around to an excited band above the conduction band
and the evanescent electron bands near $X$
wrap around to a hole band near $X$.
The surprising thing is that
in the simple analytical tunneling model,
the smaller evanescent wavevector
should result in a decay length {\em larger}
than the effective mass model when in fact
the current calculations show the opposite.
This apparent contradiction is not understood.

\subsubsection{Quantum Cascade Lasers}
\begin{figure}[h]
\begin{center}
\resizebox{3.5in}{!}{ \includegraphics{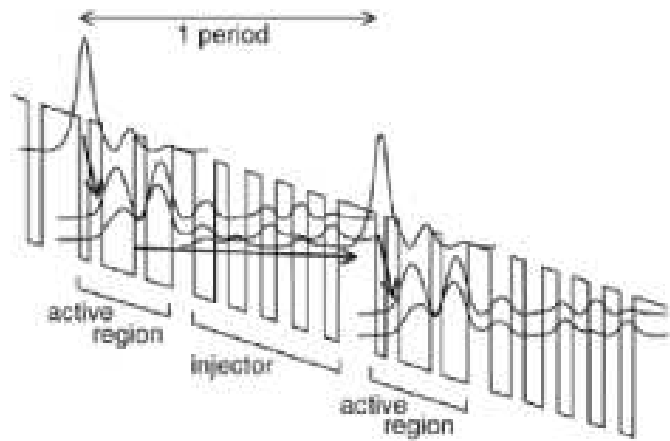} }
\end{center}
%\caption{\mbox{  }}
\caption{Quantum cascade laser structure modeled in \cite{Lee_Wacker02}. Reprinted with permission from \cite{Lee_Wacker02}. Copyright 2002 American Physical Society.}
\label{fig:Wacker_structure}
\end{figure}

The most comprehensive application of NEGF theory
to superlattices and quantum cascade lasers
was presented in Lee and Wacker \cite{Lee_Wacker02}
which built upon previous
work \cite{Wacker_PRL99,Wacker_prb02,Wacker_Phys_Reports02}.
The difficulty
with modeling quantum cascade lasers is that they are
very long containing 40 or more
periods of active and injector regions
under large bias
and a treatment of scattering
is essential since the transport through
the structure is all scattering assisted.
Lee and Wacker reformulated NEGF theory to model
these types of structures
as shown in Fig. \ref{fig:Wacker_structure}.
Their formulation of NEGF theory
takes a radical departure from all of the
other work discussed in this chapter \cite{Lee_Wacker02}.
All of the other work is based on a formulation
of NEGF theory that has the work of
Caroli et al. \cite{Caroli_I} as its foundation.
In particular all of the other work
partitions the infinite system into
a ``device''
which is the system of interest
and contacts which act as thermal reservoirs for electrons
and the partitioning is done as first described in \cite{Caroli_I}.
In the formulation of NEGF theory by Lee and Wacker,
there are no contact regions.
Special ``periodic'' boundary conditions are used
in which the Green functions and self-energies
satisfy
\begin{equation}
G_{i', j'}(E-qV) = G_{i, j}(E)
\label{eq:LeeWacker_periodicBCs}
\end{equation}
In Eq. (\ref{eq:LeeWacker_periodicBCs}),
$i$ and $j$ are the localized Wannier function
index within one period
of the structure,
$E$ is the total energy,
and
$V$ is the voltage drop across one
period of the structure.
Everything is periodic but shifted down in energy
by the voltage drop across one period.

Since there are no contacts in this theory,
an expression for current density cannot be
obtained by considering current flow
from the contact to the device.
Instead, an expression for the current density
is obtained by evaluating
\begin{equation}
J(t) = \frac{e}{{\cal V}} \langle \frac{d \hat{z}}{dt} \rangle
=
\frac{e}{{\cal V}} \frac{i}{\hbar}
\langle \left[ \hat{H} , \hat{z} \right] \rangle
\label{eq:Wacker_Jdef}
\end{equation}
The Hamiltonian is then broken up into an unperturbed part $H_0$
and a scattering part $H'$.
The unperturbed part gives rise to the standard
current expression
\begin{equation}
J_0 =
\frac{e}{\hbar} \frac{1}{A} \sum_{\kv}
\int \frac{dE}{2\pi}
{\rm tr}
\left\{
\bt_{i,i+1} \bGl_{i+1,i}
-
\bt_{i+1,i} \bGl_{i,i+1}
\right\}
\label{J0_Wacker}
\end{equation}
where the matrices $\bt$ and $\bGl$ are in the basis of
Wannier fuctions, the trace is over all Wannier functions
and spin within
a period, $\bt$ is the off-diagonal Hamiltonian matrix
coupling Wannier functions between periods $i$ and $i+1$,
$\kv$ is
the transverse wavevector, and $A$ is the cross-sectional area.
We have used the fact that $\hat{z}$ is diagonal
in the Wannier basis.
This equation is equivalent in form to Eq. (18) and (22)
of Caroli et al. \cite{Caroli_I} derived from the electron
continuity equation, Eq. (20) of Lake et al. \cite{LakeNemoTheory},
Eq. (20) of Svizhenko and Anantram \cite{Svizhenko_TED03},
and Eq. (\ref{J_general}) derived later in this chapter.

What is particularly interesting is that the scattering component
of the Hamiltonian in Eq. (\ref{eq:Wacker_Jdef}) results in a current term
\begin{equation}
J_{{\rm scatt}}
=
\frac{e}{\hbar \cal{V}}
\sum_{\kv}
\int \frac{dE}{2 \pi}
{\rm tr}
\left\{
{\bf z}
\left[ \bGl \bSA + \bGR \bSl - \bSl \bGA - \bSR \bGl \right]
\right\}
\label{eq:Jscatt}
\end{equation}
This is the first time of which we are aware that such a term
has been derived.
In a basis in which ${\bf z}$ is diagonal,
for example the Wannier basis or the
empirical tight binding basis
analyzed by Boykin \cite{Boykin_eiqr},
Eq. (\ref{eq:Jscatt}) reduces to
\begin{equation}
J_{{\rm scatt}}
=
\frac{e}{\hbar A}
\sum_{\kv}
\int \frac{dE}{2 \pi}
{\rm tr}
\left\{
\bGl \bSA + \bGR \bSl - \bSl \bGA - \bSR \bGl
\right\}
\label{eq:Jscatt_zdiag}
\end{equation}
Under very general conditions, it can be shown that
this term is equal to zero. See for example Sec. 2.4
of \cite{MahanArticle} or Eq. (67) of \cite{LakeNemoTheory}.

\section{Two Dimensional Transport in Transistors}
The primary objective for the development of quantum semiconductor
device simulators in which the potential varies in two dimensions
(2D) is the modeling of ultra-scaled Si based
FETs \cite{Jovanov_iwce00,Svizhenko_JAP02,Dresden02,Svizhenko_TED03}.
As the device size and dimension increase, computational intensity
increases, and the sophistication of the model must decrease. The
2D simulators treat the bandstructure in a multiple
single-band\index{single band} model which includes 3 decoupled
bands to account for the three non-equivalent valleys in the
quantized channel for NMOS. The simulators have included
Hartree\index{Hartree} quantum charge self-consistency, and open
system boundary conditions at the source\index{source},
drain\index{drain}, and gate\index{gate}.
For full 2D quantum transport,
empirical, elastic,
incoherent scattering\index{incoherent scattering} models have been
implemented \cite{Dresden02,Venugopal_JAP_5_03}.
The 2D
problem has been further simplified by using mode-space
models \cite{Jovanovic_JAP02,Venugopal_JAP_5_03}.

True inelastic scattering has been included
in a multiple subband model in which the subbands are only
coupled through the phonon scattering \cite{Svizhenko_TED03}.
Thus, the quantum calculations were reduced
to multiple 1D calculations greatly reducing the
computational burden.
This model was used to simulate the energy relaxation
of electrons from dispersionless deformation potential
optical phonon scattering in Si
G-type
Intervalley scattering was included
from phonons with energies of 12, 19, and 62 meV,
The electron phonon interaction was
treated in the self-consistent Born approximation,
and the resulting quantum charge density was
iterated self-consistently with a 2D Poisson's equation
to convergence.

\begin{figure}[h]
\begin{center}
\resizebox{3.5in}{!}{ \includegraphics{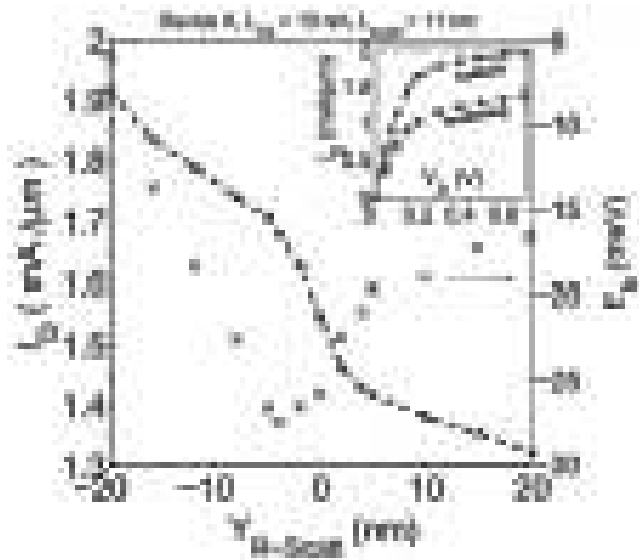} }
\end{center}
%\caption{\mbox{  }}
\caption{Drain current and channel barrier height versus
extent of scattering region (starting from
source at -20 nm).
Inset: I$_D$ vs. V$_DS$ curves for V$_G = 0.6$ V. Reprinted with permission from \cite{Svizhenko_TED03}. Copyright 2003 IEEE.}
\label{fig:Svizhenko_scattering_Y}
\end{figure}

\begin{figure}[h]
\begin{center}
\resizebox{3.5in}{!}{ \includegraphics{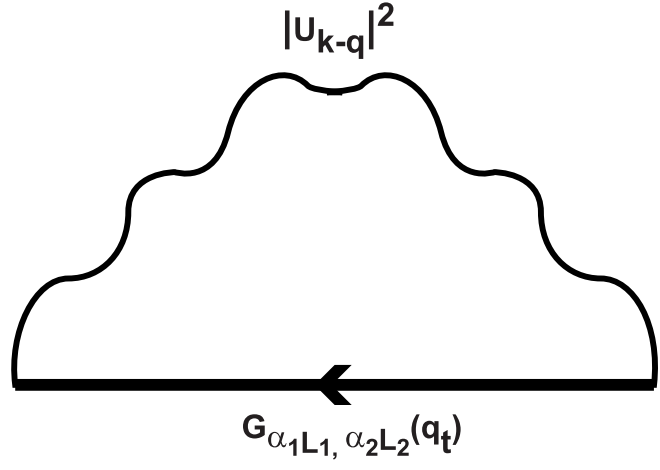} }
\end{center}
%\caption{\mbox{  }}
\caption{Self-energy diagram for self-consistent Born approximation.}
\label{fig:rainbow}
\end{figure}

Dispersionless, deformation potential optical phonon
scattering results in a
self-energy that is local in position.
In ref. \cite{Svizhenko_TED03},
the spatial scattering region was increased
starting initially in the source, then moving through
the extension and channel, and finally into the
drain to encompass the entire FET.
The effect of inelastic scattering in various regions
of the device is shown in Fig. (\ref{fig:Svizhenko_scattering_Y}).
The particular dual gate device modeled for this figure
has a 10 nm channel, 1.5 nm channel thickness, and
1.5 nm oxide.
The gate region lies between $\pm5$ nm.
As the scattering region is extended through the device
from source to drain, the effect on the current is almost
symmetric around the center of the channel.
This indicates that for ultra-scaled FETs,
scattering in the drain end is just as important as
scattering in the source end.
This contradicts the conclusion of Lundstrom and Ren
that ``scattering in a short region near the
beginning of the channel limits the on-current
\cite{Lundstro_essentialPhys02}.''
The reason for the importance of drain-end scattering
in the 10 nm FET is that the gate length is
comparable to the scattering length so that hot electrons
from the drain-end can be reflected back into the channel.
As far as we are aware, this calculation was the first to
predict these results.

The largest scale simulations of a 2D FET have been performed by
Jovanovic \cite{Dresden02}.
He performed full 2D simulations of both the MIT 90 nm and 25 nm
bulk MOSFETs \cite{Dresden02,Well_tempered_FET}.
An elastic, empirical scattering model was used and calibrated to
mobility data.
For the 90 nm device, NEGF simulations,
drift diffusion simulations, and experimental data
were all essentially indistinguishable.
However, the drift-diffusion and NEGF calculations
diverge significantly for the 25 nm device.
The differences arise from the threshold voltage shift due to the
shape of the quantum charge distribution in the channel
and tunneling through the gate region.

\subsubsection{Recursive Green Function Algorithm for $G^<$}
One of the significant algorithmic developments of the 2D codes
was the use of the recursive Green function algorithm (RGFA) for
$G^<$ \cite{Svizhenko_JAP02,Dresden02}.
The algorithm was first
developed during the NEMO program \cite{G<RGF_Jovanovic} and
implemented in early prototypes of the NEMO code, but it did not
make it into the final deliverable; only the recursive Green
fuction algorithm for $G^R$ was used \cite{LakeNemoTheory}.
In ref. \cite{Svizhenko_JAP02} the algorithm was used for coherent
transport.
In ref. \cite{Dresden02}, the algorithm was used for transport
in the presence of incoherent scattering.
This was the first use of which we are aware of a recursive Green
function algorithm used in the presence of incoherent
scattering.
It was developed to meet the intense computational demands
of modeling the large 90nm and 25nm MIT bulk MOSFET devices.
Below we give an overview of the algorithm.

First, we note that
for elastic incoherent scattering, $G^<$ can be factored into
components injected from the left and right contacts.
This can be seen from Eqs. (73) - (76) of ref. \cite{LakeNemoTheory}
or Eq. (13) of \cite{Dresden02}.
The empirical scattering model has a local, diagonal form for the
self energies,
\begin{equation}
\Sigma^R_{i,i} = -i \Gamma_{i,i} / 2
\label{SigmaR_empirical}
\end{equation}
\begin{equation}
\Sigma^<_{i,i} = \frac{\Gamma_{i,i}}{A_{i,i}} G^<_{i,i}
\label{Sigma<_empirical}
\end{equation}
The form for $\Sigma^<$ conserves current.
The general form for $G^<$ is
\begin{equation}
G^< = G^R \Sigma^< G^A + G^R \Sigma^{<B} G^A
\label{G<general}
\end{equation}
The first term on the right
is from incoherent scattering, and
the second term on the right is a source term due to injection from
the contacts.
The starting point for the RGFA for $G^<$ is the equation
for the left connected $g^<$
\begin{equation}
\gll_{i,i} = \grl_{i,i} \left( \mbox{$\sigma$}^s_{i,i} + t_{i,i-1} \gll_{i-1,i-1} t_{i-1,i} \right)
\gal_{i,i}
\label{rgf_gl}
\end{equation}
The superscripts $\lhd$ and $\rhd$ are used to indicate whether
the semi-infinite portion of $g^<$, $g^R$, $g^A$, or $a^s$ extends towards
the left (source) or right (drain) side, respectively.
$\mbox{$\sigma$}^s_{i,i} = \frac{\Gamma_{i,i}}{A_{i,i}} G^{<s}_{i,i}$
and
$G^{<s}_{i,i} = i f^s A^s_{i,i}$ where the superscript $s$ indicates
source.
Writing the equivalent equation for $\ggl$ and subtracting gives an equation for the left connected
source spectral function\index{spectral function}
\begin{equation}
\label{eq:silsc}
{a}^{\lhd s}_{i,i} = {g}^{\lhd r}_{i,i} \left[ \frac{{\Gamma}_{i,i}}{{A}_{i,i}} {A}^s_{i,i} + {t}_{i,i-1} {a}^{s}_{i-1,i-1} {t}_{i-1,i} \right] {{g}}^{\lhd a}_{i,i}
\end{equation}
and, similarly, the {\em right} connected {\em source} spectral function\index{spectral function}
\begin{equation}
\label{eq:sirsc}
{a}^{\rhd s}_{i,i} = {g}^{\rhd r}_{i,i} \left[ \frac{{\Gamma}_{i,i}}{{A}_{i,i}} {A}^s_{i,i} + {t}_{i,i+1} {a}^{\rhd s}_{i+1,i+1} {t}_{i+1,i} \right] {{g}}^{\rhd a}_{i,i}.
\end{equation}
The left connected source spectral function is connected to the source
contact. The {\em right} connected source spectral function
is not connected to either the source or the drain. It is only non-zero
if there is incoherent scattering, i.e. if $\Gamma$ is non-zero.
With the right and left connected source spectral functions,
the full source spectral function is calculated from
\begin{equation}
\label{eq:Gltra}
{A}^s_{i,i} = {G}^r_{i,i} \left[ \frac{{\Gamma}_{i,i}}{{A}_{i,i}} {A}^s_{i,i} + {t}_{i,i-1} {a}^{\lhd s}_{i-1,i-1} {t}_{i-1,i} + {t}_{i,i+1} {a}^{\rhd s}_{i+1,i+1} {t}_{i+1,i} \right] {G}^{a}_{i,i}.
\end{equation}
The above 3 equations must be iterated until convergence at which point
current is conserved.

\section{Three Dimensional Transport through CNTs, Nanowires, and Molecules}
Three dimensional (3D) modeling
has been applied to semiconductor nanowires, carbon nanotubes, and molecules.
In all of the 3D applications of which we are aware, the transport
is modeled as coherent throughout the nanowire, CNT, or molecule.
This may be a good approximation in CNTs, but for electrons
traveling through the HOMO or LUMO bands of molecules,
the approximation is open to question.
The 3D modeling
can be roughly classified into
two categories. In the first category, the Hamiltonian is constructed
using an empirical tight binding model. For well-understood
materials such as the common semiconductors, Si, Ge, and GaAs, and
carbon nanotubes (CNTs),
the empirical models can be highly accurate since they are fitted to
known properties of the materials such as band gaps and effective
masses \cite{Jancu,Boykin_04}. In fact, for semiconductors, the empirical
models are more accurate than the {\em ab-initio} models.
Carbon nanotubes
have been extensively modeled using the
$\pi$-bond model \cite{Anant_CNTdisorder98,Anant_CNTcaps00,J_Guo_Scaling_CNTs04,Register_JAP04}.
In the $\pi$-bond model, the basis consists of one $p_z$ orbital
per atom perpendicular to the axis of the nanotube \cite{Saito_CNTbook}.
The 3D CNT problem has also been simplified by assuming cylindrical symmetry
and using a mode-space model \cite{J_Guo_Scaling_CNTs04}.
Applications of NEGF to 3D modeling of semiconductors have been
more scarce.
Both a full-band sp$^3$d$^5$s$^*$ model and a discretized single band effective mass model
were used to calculate transmission
coefficients through Si nanowires \cite{Y-J_Ko,Nanotech03,Rivas_Lake}.
No attempt has yet been made to combine the NEGF calculation with
a self-consistent Poisson solver and calculate I-V characteristics of
the Si nanowires under bias.

In the second category, the Hamiltonian is constructed
using {\em ab initio} models with density functional theory (DFT)
and a localized orbital basis \cite{Taylor_ab_init_PRB01,Seminario_NEGF01,Xue_JChemPhys01,Xue_ChemPhys02,Ghosh_Datta_PRB01,Taylor_Mag_CNT_PRL00,Taylor_Res_CNT_PRB01,Taylor_C60_PRB01,Taylor_IV_N_CNT_PRB02,Anant_PRL02,Y_Xue_thesis,Verges_Nanotech02,Schon_PRL02,Verges_PRB03,Xue_PRB03,Seminario_ProcIEEE03,Seminario_program.diode03,Xue_PRB04,Ghosh_gating04}.
In this category, NEGF has been integrated with existing codes such as
the quantum chemistry code Gaussian98 \cite{Seminario_NEGF01,Xue_JChemPhys01,Ghosh_Datta_PRB01,Xue_ChemPhys02},
the computational materials codes SIESTA \cite{Brandbyge02,Stokbro_ANYAS03}
and FIREBALL \cite{Demkov_PhysStatSol02},
and custom codes \cite{Taylor_ab_init_PRB01,Schon_PRL02}.
It is in this last, highly interdisciplinary category that we find
some of the most interesting recent developments.

The general field of molecular electronics has brought together quantum chemists,
physicists, material scientists, and electrical engineers.
The two primary fields from which theoretical approaches have been taken are
solid state physics and computational quantum chemistry.
Diagrammatic perturbation theory, Green's functions,
and Dyson's equation are the foundation of
much of modern solid state physics theory, and it is the basis for
non-equilibrium Green
function theory \cite{KadanoffBaym,FetterWalecka,MAHANBOOK}.
Diagrammatic perturbation theory has had less usage
in modern quantum chemistry \cite{SzaboOstlund,PropQuChem}.
Conversely, non-orthogonal basis sets are often used in
chemistry \cite{non_ortho_basis_bridge,Effect_H_nonortho_basis03},
but are generally only considered in physics
by those working on computational material properties using a localized
orbital basis \cite{Lannoo,WilliamsLang82,rgf_volume,Harrison2,Fulde,TBinding_MRS,Inglesfield01}.
One exception is Feynman's derivation of the anticommutation properties
of creation and annihilation operators in a
non-orthogonal basis \cite{Feynman_Stat_mech}.
The derivations in the literature describing the integration of
DFT with NEGF are extensive \cite{Seminario_NEGF01,Taylor_ab_init_PRB01,Xue_JChemPhys01,Ghosh_Datta_PRB01,Brandbyge02,Xue_ChemPhys02},
although there has been little discussion of the validity of
DFT in non-equilibrium \cite{Y_Xue_thesis,DFT_validity_PRB04}.
Recent work suggests that the breakdown of this approximation
may be the reason for the often observed
large discrepancy between theoretical
and experimental current magnitudes \cite{DFT_validity_PRB04}.
Also, the derivations make little use of second quantized operators
which are the natural language of Green function and diagramatic perturbation
theory \cite{KadanoffBaym,FetterWalecka,MAHANBOOK}.

Below, we provide tutorial level derivations of
standard expressions starting from their basic definitions
in second quantized form in a non-orthogonal basis for an effective
single particle Hamiltonian.
The focus will be on the new problems that arise when
working in a non-orthogonal basis as compared to an orthogonal
basis.
We will point out open questions and
standard approximations of which
one finds little discussion in the literature.

\section{Green Functions in a Non-orthogonal basis}
\subsection{The Non-Orthogonal Localized Orbital Basis}
We start with a non-orthogonal localized orbital basis $\{ \ket{ \alp_i } \}$
where the states correspond
to the wavefunctions $\phi_i(\rv) = \braket{\rv}{\alp_i}$.
We next define the biorthogonal basis $\{ \ket{ \bb_i } \}$
such that
\begin{equation}
\braket{ \alp_i }{\bb_j} = \delta_{i,j}
\label{biorthogonal_prop}
\end{equation}
The overlap matrix is defined as
$S_{i,j} = \braket{\alp_i}{\alp_j}$.
By inspection, the orbitals $\{ \ket{ \bb_j } \}$ are written as linear
combinations of the $\{ \ket{ \alp_i } \}$ as
\begin{equation}
\ket{\bb_j} = \sum_p \ket{\alp_p} [S^{-1}]_{p,j}
\label{beta_i_construct}
\end{equation}
since this satisfies Eq. (\ref{biorthogonal_prop}).
With the expansion (\ref{beta_i_construct}),
we can calculate the inner product
\begin{equation}
\braket{\bb_i}{\bb_j} =
\sum_q \left[ S^{-1} \right]^{\dagger}_{i,q}
\underbrace{
\braket{\alp_q}{\alp_p} \left[ S^{-1} \right]_{p,j}
}_{ \delta_{q,j} }
\nonumber
\\
=
\left[ S^{-1} \right]^{\dagger}_{i,j} \; = \; \left[ S^{-1} \right]_{i,j}
\label{bi_bj}
\end{equation}
The last equality results from the fact that the inverse of a Hermitian
matrix is also Hermitian.

In our non-orthogonal basis $\{ \ket{ \alp_i } \}$, the identity operator is
\begin{equation}
\Iop = \sum_{i,j}\ket{\alp_i} \left[ S^{-1} \right]_{i,j} \bra{\alp_j}
=
\sum_j \ket{\bb_j} \bra{\alp_j} = \sum_j \ket{\alp_j} \bra{\bb_j}
\label{I_op}
\end{equation}
This is proven by sandwiching the identity operator between any two
basis states,
\begin{equation}
\braket{a_k}{a_l}=
\sum_{i,j}
\underbrace{ \braket{a_k}{a_i} }_{\left[ S \right]_{k,i}}
\left[ S^{-1} \right]_{i,j}
\underbrace{ \braket{\alp_j}{a_l} }_{\left[ S \right]_{j,l}}
\nonumber
\\
=
\left[ S \right]_{k,l}
\end{equation}

We next define the creation, $a_i^{\dagger}$, and annihilation, $a_i$,
operators for the non-orthogonal
basis  states $\{ \ket{ \alp_i } \}$ in the usual way such that
$a^{\dagger}_i$ creates a particle in the empty state $\ket{\alp_i}$.
In other words, starting from
the vacuum state, $\ket{0}$, the single particle state
$\ket{a_i}$ is filled by applying the creation operator,
\begin{equation}
a^{\dagger}_i \ket{0} = \ket{\alp_i}.
\label{a_dag_0}
\end{equation}
The anti-commutation relations are \cite{Feynman_Stat_mech}
\begin{equation}
\left\{ a_i, a_j^{\dagger} \right\}=S_{i,j}.
\label{a_anticomm_relation}
\end{equation}
and
$\left\{ a_i, a_j \right\} = \left\{ a_i^{\dagger}, a_j^{\dagger} \right\} = 0$.
We also define the creation, $b^{\dagger}$, and annihilation, $b$,
operators for the bi-orthogonal
basis  states $\{ \ket{ \bb_i } \}$ such that
\begin{equation}
b^{\dagger}_i \ket{0} = \ket{\bb_i}.
\label{b_dag_0}
\end{equation}
Substituting the expansion of state $\ket{\bb_i}$
in terms of states $\ket{\alp_j}$, Eq. (\ref{beta_i_construct}),
into the right hand side of Eq. (\ref{b_dag_0}),
and then substituting in Eq. (\ref{a_dag_0}) for
$\ket{\alp_j}$, we obtain the transformations
between the 2 sets of operators,
\begin{eqnarray}
b_j^{\dagger}&=&\sum_p a_p^{\dagger} \left[ S^{-1} \right]_{p,j}
\nonumber
\\
b_j&=&\sum_p \left[ S^{-1} \right]_{j,p} a_p
\label{a_b_transforms}
\end{eqnarray}
With these expansions and Eq. (\ref{a_anticomm_relation}),
it is straightforward to calculate the anticommutation
relations
\begin{eqnarray}
\left\{ b_i, b^{\dagger}_j \right\}&=&\left[ S^{-1} \right]_{i,j}
\\
\label{b_anticomm_relation}
\left\{ a_i, b^{\dagger}_j \right\}&=&\left\{ a_i^{\dagger}, b_j \right\} =
\delta_{i,j}
\label{mixed_anticomm_relation}
\end{eqnarray}
and $\{b_i, b_j\} = \{b^{\dagger}_i, b^{\dagger}_j\} =
\{a_i, b_j\} = \{a^{\dagger}_i, b^{\dagger}_j\} = 0$.

We also introduce the standard field operators,
$\psi^{\dagger}(\rv)$ and $\psi(\rv)$,
which are the creation and annihilation operators corresponding
to the orthonormal position eigenstates $\ket{\rv}$.
A particle in state $\ket{\rv}$ is created by
\begin{equation}
\psi^{\dagger}(\rv) \ket{0} = \ket{\rv}.
\label{psi_dag_0}
\end{equation}
These have the standard anti-commutation relations,
$\{ \psi^{\dagger}(\rv '), \psi(\rv) \} = \delta(\rv - \rv ')$.
By applying the identity operator, Eq. (\ref{I_op}), to the right hand side of
Eq. (\ref{psi_dag_0}),
\begin{eqnarray}
\ket{\rv}&=&\sum_{i,j}\ket{\alp_i} \left[ S^{-1} \right]_{i,j} \braket{\alp_j}{\rv}
\nonumber
\\
&=&
\sum_{i,j}\ket{\alp_i} \left[ S^{-1} \right]_{i,j} \phi^*_j(\rv)
\nonumber
\\
&=&\sum_{j}\ket{\bb_j} \phi^*_j(\rv)
\label{r_expansion}
\end{eqnarray}
and then expanding the state $\ket{\alp_i}$ using
Eq. (\ref{a_dag_0}), we obtain the field operators in terms of the
localized orbital operators,
\begin{eqnarray}
\psi^{\dagger}(\rv)&=&\sum_{i,j} a_i^{\dagger} \left[ S^{-1} \right]_{i,j} \phi^*_j(\rv)
\nonumber
\\
&=& \sum_j b_j^{\dagger} \phi^*_j(\rv)
\label{psi_dag_expansion}
\end{eqnarray}
Similarly
\begin{equation}
\psi(\rv) = \sum_j \phi_j(\rv) b_j
\label{psi_expansion}
\end{equation}

It is useful to introduce the 3 different kinds of matrix elements
of any operator \cite{Lannoo}.
We will use the identity operator as an example.
The covariant matrix elements are
\begin{equation}
\bra{\alp_i} \Iop \ket{\alp_j}\;=\;\left[S\right]_{i,j} \; \equiv \; I_{i,j}.
\end{equation}
In other words, the overlap matrix is the covariant representation
of the identity operator.
The contravariant matrix elements are
\begin{equation}
\bra{\bb_i} \Iop \ket{\bb_j}\;=\;\left[S^{-1}\right]_{i,j} \; \equiv \; I^{i,j}.
\end{equation}
The inverse of the overlap matrix is the contravariant representation
of the identity operator.
The mixed representation of the identity operator is the usual identity
matrix,
\begin{equation}
\bra{\bb_i} \Iop \ket{a_j}\;=\; \delta_{i,j} \; \equiv \; I^i_{\: j} \:
\end{equation}
where $\delta_{i,j}$ is the usual kronecker delta function.
These representations of the identity operator act like the metric
tensor and serve to raise and lower indices, e.g.,
\begin{equation}
\sum_j I_{i,j} I^{j,k} = I_i^{\, k} .
\end{equation}

From Eqs. (\ref{beta_i_construct}) and (\ref{a_b_transforms}), we see that
the states
$\{ \ket{ \bb_j } \}$ and the creation operators $b_j^{\dagger}$
transform in the same way as contravariant quantities.
If we were to use superscripts and subscripts to differentiate between the
covariant and contravariant states,
we could re-write Eq. (\ref{beta_i_construct}) as
\begin{equation}
\ket{\alp^j} \equiv \ket{\bb_j} = \sum_p \ket{\alp_p} I^{p,j}
\label{beta_i_construct_contrav}
\end{equation}
From Eqs. (\ref{psi_dag_expansion}) and (\ref{psi_expansion}), we see
that the field operators are contractions of a covariant and contravariant
quantity and are thus independent of the localized orbital basis
as they must be.
We find this to be a useful check when performing derivations
that any operator corresponding to an observable is
a fully contracted product of covariant and contravariant quantities.

\subsection{Non-equilibrium Green functions and correlation functions}
We will now derive expressions for the electron density and the
current in terms of the relevant Green function quantities.
Two correlation functions, 2 Green functions,
and 1 spectral function are of particular interest.
The 2 correlation functions are defined by
\begin{eqnarray}
G^<_{i,j}(t,t')&=&\frac{i}{\hbar} \langle b^{\dagger}_j(t') b_i(t) \rangle
\label{G<}
\\
G^>_{i,j}(t,t')&=&\frac{-i}{\hbar} \langle b_i(t) b^{\dagger}_j(t') \rangle .
\label{G>}
\end{eqnarray}
The retarded and advanced Green functions are defined by
\begin{eqnarray}
G^R_{i,j}(t,t')&=&\Theta(t-t')\frac{-i}{\hbar}
\langle b_i(t) b^{\dagger}_j(t') +  b^{\dagger}_j(t') b_i(t) \rangle
\nonumber \\
&=& \Theta(t-t') \left[ G^>_{i,j}(t,t') - G^<_{i,j}(t,t') \right]
\label{GR}
\\
G^A_{i,j}(t,t')&=&\Theta(t'-t)\frac{i}{\hbar}
\langle b_i(t) b^{\dagger}_j(t') +  b^{\dagger}_j(t') b_i(t) \rangle .
\label{GA}
\end{eqnarray}
The spectral function is defined by
\begin{eqnarray}
A_{i,j}(t,t')&=&\frac{1}{\hbar}\langle  b_i(t) b^{\dagger}_j(t') +  b^{\dagger}_j(t') b_i(t) \rangle
\\
&=& i \left[ G^R_{i,j}(t,t') - G^A_{i,j}(t,t') \right]
\\
&=& i \left[ G^>_{i,j}(t,t') - G^<_{i,j}(t,t') \right]
\end{eqnarray}
where the brackets $\langle \cdots \rangle$
indicate the non-equilibrium ensemble
average \cite{Craig,Langreth,Selman_negf},
and
the creation and annihilation operators are in the Heisenberg
representation.
We will be concerned with steady state in which case the time
dependence becomes only a function of $(t-t')$ and
is Fourier transformed to energy, e.g.
\begin{equation}
G^<_{i,j}(E) = \int d(t-t') e^{iE(t-t')/\hbar} G^<_{i,j}(t,t')
\label{Fourier}
\end{equation}

Note that these Green functions, being products of two
contravariant quantities, are the contravariant representation
of the Green function. It is this representation which is most
useful for numerical computations.
We will see that
$G^R_{i,j}(E)$ is given by the inverse of
the matrix whose elements are the covariant representation
of $[E-H]$, i.e. the inverse of the matrix whose
elements are $\bra{\alp_i} E-H \ket{\alp_j} = ES_{i,j} - H_{i,j}$.
This matrix is sparse and therefore desirable for numerical
manipulation.

The expression for the electron density is straightforward
to obtain,
and it is directly related to $G^<$.
\begin{eqnarray}
\langle \hat{n}(\rv) \rangle &=&\langle \psi^{\dagger}(\rv) \psi(\rv) \rangle
\nonumber
\\
&=&\sum_{i,j} \langle b^{\dagger}_j b_i \rangle \phi_j^*(\rv) \phi_i(\rv)
\nonumber
\\
&=& -i \hbar \sum_{i,j} G^<_{i,j}(t=t'=0) \phi_j^*(\rv) \phi_i(\rv)
\nonumber
\\
&=& -i \sum_{i,j}
\int \frac{dE}{2 \pi} G^<_{i,j}(E) \phi_j^*(\rv) \phi_i(\rv)
\label{n_def}
\end{eqnarray}
The expression for the current is less straightforward and we
will defer its derivation until we have the equations
of motion for $G^<$ and $G^R$.

The equations of motion for the Green functions are derived by applying
$i \hbar \partial / \partial t$ to expressions (\ref{G<}) - (\ref{GR})
and evaluating.
\begin{equation}
i \hbar \frac{\partial}{\partial t} \langle b^{\dagger}_j(t') b_i(t) \rangle =
\langle b^{\dagger}_j(t') \left[b_i, \hat{H} \right] \rangle
\label{eq_motion1}
\end{equation}
where the effective single particle Hamiltonian is
\begin{equation}
\hat{H} = \sum_{i,j} H_{i,j} b_i^{\dagger} b_j .
\label{H}
\end{equation}
The commutator in (\ref{eq_motion1}) is evaluated using
the anticommutation relation
Eq. (\ref{b_anticomm_relation}).
\begin{equation}
\left[b_i, \hat{H} \right] =
\sum_{k,l} H_{k,l} \left[b_i, b_k^{\dagger} b_l \right]
=
\sum_{k,l} \left[ S^{-1} \right]_{i,k} H_{k,l} b_l(t)
\label{eq_motion_comm}
\end{equation}
Placing this back into Eq. (\ref{eq_motion1}) gives
\begin{equation}
i \hbar \frac{\partial}{\partial t} G^<_{i,j}(t,t') =
\sum_{k,l} \left[ S^{-1} \right]_{i,k} H_{k,l}
G^<_{l,j}(t,t')
\label{G<_eom}
\end{equation}
Going through the same exercise for $G^>_{i,j}(t,t')$ results in the
same equation of motion as for $G^<_{i,j}(t,t')$.
Applying $i \hbar \partial / \partial t$
to the retarded Green function, Eq. (\ref{GR}), gives
\begin{equation}
i \hbar \frac{\partial}{\partial t} G^R_{i,j}(t,t') =
\delta(t-t') \left[ S^{-1} \right]_{i,j}
+
\sum_{k,l} \left[ S^{-1} \right]_{i,k} H_{k,l} G^R_{l,j}(t,t')
\label{GR_eom1}
\end{equation}
Multiplying through by $S$ gives
\begin{equation}
i \hbar \sum_l S_{i,l} \frac{\partial}{\partial t} G^R_{l,j}(t,t') =
\delta(t-t') \delta_{i,j}
+
\sum_l H_{i,l} G^R_{l,j}(t,t')
\label{GR_eom2}
\end{equation}
or in matrix notation
\begin{equation}
\left[
i \hbar {\bf S} \frac{\partial}{\partial t} - {\bf H} \right]
{\bf G^R}(t,t')
=
{\bf 1} \delta(t-t')
\label{GR_tt'}
\end{equation}
Fourier transforming to energy as in Eq. (\ref{Fourier}) results in
\begin{equation}
\left[
E {\bf S}  - {\bf H} \right]
{\bf G^R}(E)
=
{\bf 1}
\label{GR_E}
\end{equation}
Similarly, Eq. (\ref{G<_eom}) becomes
\begin{equation}
\left[
E {\bf S}  - {\bf H} \right]
{\bf G^<}(E)
=
0
\label{G<_E}
\end{equation}

\subsection{Boundary self-energies}
\subsubsection{$\Sigma^R$ and $G^R$}
Now one partitions the infinite system into a finite ``device'' region
and ``contact''
regions \cite{Caroli_I,Meir_Wingreen,DattaTextbook2,LakeNemoTheory}.
Traditionally,
one then includes the effect of the ``contacts'' on the
``device'' exactly using time dependent diagrammatic perturbation theory
summed to all orders to obtain the Dyson's equations for each type of
Green function \cite{Caroli_I,Meir_Wingreen,LakeNemoTheory}.
Time dependent perturbation theory is for perturbations in the
Hamiltonian which cause the states to
evolve according to unitary transformations.
We will see that now we need to also include a perturbation in
the off diagonal elements of the overlap matrix. This, in effect,
alters the basis states and thus the Hilbert space spanned by the states.
%so that an initial state and a final state do not necessarily have the
%same norm.
Therefore, it is not clear to us how to incorporate this into
the formal nonequilibrium Green's function theory.
What we will do is what many others have
done \cite{Lannoo,DattaTextbook2,Seminario_NEGF01,Taylor_ab_init_PRB01,Xue_JChemPhys01,Ghosh_Datta_PRB01,Brandbyge02,Xue_ChemPhys02},
which is to
use matrix algebra to write an effective Dyson equation for
$G^R$ to include the effect of the contacts.
This is an application of inversion by partitioning \cite{N_RECIPIES_I_P}.

We illustrate the concept with a concrete example of a periodic chain of atoms
with nearest neighbor overlap of the atomic-orbital-like basis states.
Atoms $\{-\infty, \ldots , 0\}$ lie in the left contact. Atoms $\{1, \ldots,  N\}$ lie in the
``device'' and atoms $\{(N+1), \ldots , \infty\}$ lie in the right contact.
The matrix elements of the Hamiltonian group into intra-atomic subblocks
${\bf D}_{i,i}$ and inter-atomic subblocks ${\bf t}_{i,i \pm 1}$.
The size of these matrices is equal to the number of orbitals
per atom.
In the matrix $[E{\bf S} - {\bf H}]$, the matrix elements of the
left contact
couple to those of the device by the off-diagonal blocks
\begin{equation}
E{\bf S}_{1,0} - {\bf t}_{1,0} \doteq  - \tilde{\bf t}_{1,0}
\label{t10_tilde}
\end{equation}
and
\begin{equation}
E{\bf S}_{0,1} - {\bf t}_{0,1} \doteq  - \tilde{\bf t}_{0,1} .
\label{t01_tilde}
\end{equation}
The matrix elements of the right contact couple to
those of the device by the off-diagonal blocks
\begin{equation}
E{\bf S}_{N,N+1} - {\bf t}_{N,N+1} \doteq  - \tilde{\bf t}_{N,N+1}
\label{tN_N1_tilde}
\end{equation}
and
\begin{equation}
E{\bf S}_{N+1,N} - {\bf t}_{N+1,N} \doteq  - \tilde{\bf t}_{N+1,N} .
\label{tN1_N_tilde}
\end{equation}

For any atoms $i,j \in \{1, \ldots , N\}$ in the device,
we can write the standard Dyson equation for the
retarded Green function using
$\tilde{\bf t}$ in place of ${\bf t}$.
\begin{eqnarray}
{\bf G}^R_{i,j} =
{\bf g}^R_{i,j}&+&
{\bf g}^R_{i,1} \tilde{\bf t}_{1,0} {\bf G}^R_{0,j}
\nonumber \\
&+&
{\bf g}^R_{i,N} \tilde{\bf t}_{N,N+1} {\bf G}^R_{N+1,j}
\label{Dyson1}
\end{eqnarray}
where ${\bf g}^R$ is the ``bare'' Green function of the uncoupled device,
and the energy argument is suppressed.
We write a second Dyson equation for the exact ${\bf G}^R_{0,j}$
and ${\bf G}^R_{N+1,j}$ which cross the device-contact boundaries.
\begin{eqnarray}
{\bf G}^R_{0,j}&=&{\bf g}^R_{0,0} \tilde{\bf t}_{0,1} {\bf G}^R_{1,j}
\label{G0j}
\\
{\bf G}^R_{N+1,j}&=&{\bf g}^R_{N+1,N+1} \tilde{\bf t}_{N+1,N}
{\bf G}^R_{N,j}
\label{GNp1j}
\end{eqnarray}
where ${\bf g}^R_{0,0}$ is the ``bare''
Green function of the end block of the uncoupled left contact
(also referred to as the ``surface'' Green function), and
${\bf g}^R_{N+1,N+1}$ is the ``bare'' Green function of the
uncoupled right contact.
These are generally calculated from
the eigenmodes of the bulk contact
regions \cite{CR_JAP1,Rivas_Lake,Ghosh_Datta_PRB01,DattaTextbook2}.
Combining Eqs. (\ref{Dyson1}), (\ref{G0j}), and (\ref{GNp1j}) gives the
standard Dyson equation for the device Green function.
\begin{eqnarray}
{\bf G}^R_{i,j} = {\bf g}^R_{i,j}&+&
{\bf g}^R_{i,1}
\underbrace{ \tilde{\bf t}_{1,0}
{\bf g}^R_{0,0} \tilde{\bf t}_{0,1} }_{{\bf \Sigma}^{RB}_{1,1}}
{\bf G}^R_{1,j}
\nonumber \\
&+&
{\bf g}^R_{i,N}
\underbrace{ \tilde{\bf t}_{N,N+1}
{\bf g}^R_{N+1,N+1} \tilde{\bf t}_{N+1,N} }_{{\bf \Sigma}^{RB}_{N,N}}
{\bf G}^R_{N,j}
\label{Dyson3}
\end{eqnarray}
The retarded self-energies resulting from coupling to the left and right
contacts are, respectively,
\begin{equation}
{\bf \Sigma}^{RB}_{1,1} = \left( {\bf t}_{1,0} - E{\bf S}_{1,0} \right)
{\bf g}^R_{0,0}
 \left( {\bf t}_{0,1} - E{\bf S}_{0,1} \right)
\label{Sigma_RL}
\end{equation}
\begin{equation}
{\bf \Sigma}^{RB}_{N,N} = \left( {\bf t}_{N,N+1} - E{\bf S}_{N,N+1} \right)
{\bf g}^R_{N+1,N+1}
 \left( {\bf t}_{N+1,N} - E{\bf S}_{N+1,N} \right)
\label{Sigma_RR}
\end{equation}
We add the superscript $B$ to indicate that these self-energies
result from the open-system boundary conditions of the device,
and they are not the result of dissipative processes within the
device.
Multiplying Eq. (\ref{Dyson3}) by the $N \times N$ block matrix
of the isolated device,
$[ E{\bf S}_0 - {\bf H}_0 ] = \left[ {\bf g}^R \right]^{-1}$,
results in standard equation for the device Green function.
\begin{equation}
\left[ E{\bf S}_0 - {\bf H}_0 -  {\bf \Sigma}^{RB} \right] {\bf G}^R = {\bf 1}
\label{GR_device_eom}
\end{equation}
where ${\bf \Sigma}^{RB}$ has two nonzero blocks,
${\bf \Sigma}^{RB}_{1,1}$ and ${\bf \Sigma}^{RB}_{N,N}$.

At this point, we have derived the matrix equation for the contravariant
retarded Green function, Eq. (\ref{GR_E}),
using the Heisenberg equation of motion, Eq. (\ref{eq_motion1}),
and then we used matrix
algebra to determine the values of ${\bf G}^R_{i,j}$ in the central
device region in terms of the surface Green functions of the
contacts, Eq. (\ref{GR_device_eom}).
So far, we have not needed to know anything about time contours and
nonequilibrium Green function theory.

\subsubsection{$\Sigma^<$ and $G^<$}
To calculate physical observables such as the
electron density and current,
we need $G^<$.
To obtain $G^<$,
we need an expression for the self-energy
$\Sigma^{<B}$.
This cannot be obtained without recourse to NEGF theory.
However, as we noted above, perturbation of the basis states
does not appear to be compatible with formal NEGF theory.
We will have to choose a form for $\Sigma^{<B}$ that
appears to be consistent with the NEGF theory, that is
physically reasonable, that satisfies the usual relations
between $\Sigma^R$, $\Sigma^{<B}$, and $\Sigma^{>B}$,
and that reduces to the correct form in equilibrium.

Once ${\bf \Sigma}^{RB}$ is fixed,
Eqs. (\ref{Sigma_RL}) and (\ref{Sigma_RR}),
certain relationships between the various self
energies must be satisfied such as
\begin{equation}
\BG^B = i\left[{\bf \Sigma}^{RB} -
\left({\bf \Sigma}^{RB}\right)^{\dagger}\right]
= i \left[{\bf \Sigma}^{>B} -
{\bf \Sigma}^{<B} \right] .
\label{Sigma_sum_rule}
\end{equation}
Furthermore, in equilibrium, ${\bf \Sigma}^{<B}$ is completely
determined by ${\bf \Sigma}^{RB}$ from the relation
\begin{equation}
{\bf \Sigma}^{<B} = i f \BG^B
\label{Sigma<_equilibrium}
\end{equation}
where $f$ is the Fermi factor.
These two relations considerably limit the possible forms for
$\Sigma^{<B}$.
One also finds the argument in the literature
that since the contacts are in
equilibrium, ${\bf \Sigma}^{<B}_{1,1}$ must equal
$i f_L \BG^B_{1,1}$.
However, ${\bf \Sigma}^{<B}_{1,1}$ is
a self-energy of the ``device'' consisting of matrix
elements of atom 1 which is not in equilibrium.
Therefore, we are not convinced of the validity of this
argument even though it does give the correct answer.

$G^<$ is obtained from
the Dyson equation for $G^<$ which, in turn, is
obtained from the
Dyson equation for the contour ordered Green function by
applying the rules of Langreth \cite{Caroli_I,Langreth}.
The general form is
\begin{equation}
{\bf G}^< = {\bf g}^< + {\bf g}^R {\bf \Sigma}^R {\bf G}^<
+
{\bf g}^R {\bf \Sigma}^< {\bf G}^A
+
{\bf g}^< {\bf \Sigma}^A {\bf G}^A
\label{Dyson_G<}
\end{equation}
For a single electron
perturbation, ${\bf H'}$, the Dyson equation for
the contour ordered Green function is \cite{non-ortho_Wickes_thm}
\begin{equation}
{\bf G}^P (s,s') = {\bf g}^P(s,s') +
\int_c \! ds_1 \! \int_c \! ds_2 \: {\bf g}^P(s,s_1)
{\bf H}^{\prime} \delta(s_1 - s_2) {\bf G}^P(s_2, s')
\label{GP_1e}
\end{equation}
where the integrals are along the
Keldysh contour \cite{Langreth,Keldysh}.
Since the perturbing potential $H^{\prime}$ is local in time,
$\propto \delta(s_1 - s_2)$,
$s_1$ and $s_2$ are always on the same side of the Keldysh contour
resulting in the equation for $G^R$ and $G^<$ being,
respectively \cite{Langreth,Caroli_I},
\begin{equation}
{\bf G}^R (t,t') = {\bf g}^R(t,t') +
\int dt_1 \int dt_2 \: {\bf g}^R(t,t_1)
{\bf H}^{\prime} \delta(t_1 - t_2) {\bf G}^R(t_2, t')
\label{GR_t}
\end{equation}
and
\begin{eqnarray}
{\bf G}^< (t,t')&=&{\bf g}^<(t,t')
\nonumber \\
&+&
\int dt_1 \int dt_2 \: {\bf g}^R(t,t_1)
{\bf H}^{\prime} \delta(t_1 - t_2) {\bf G}^<(t_2, t')
\nonumber \\
&+&
\int dt_1 \int dt_2 \: {\bf g}^<(t,t_1)
{\bf H}^{\prime} \delta(t_1 - t_2) {\bf G}^A(t_2, t') .
\label{G<_t}
\end{eqnarray}
In Eq. (\ref{G<_t}), the term in Eq. (\ref{Dyson_G<}) containing the
self energy ${\bf \Sigma}^<$ is missing because the
local time dependence of the potential forces
$s_1$ and $s_2$ in (\ref{GP_1e})
to always reside on the same branch of the contour.
We now want to check that our new energy-dependent
effective potential $\tilde{\bf t} = ({\bf t} - E{\bf S})$
is still local in time and that it does not introduce any
memory effects which could give rise to a new term
in the usual Dyson equation for $G^<$.
To check that $\tilde{\bf t} = ({\bf t} - E{\bf S})$,
is still local in time,
we need to consider Eqs. (\ref{Dyson1} - \ref{GNp1j})
in the time domain.
Writing down Eq. (\ref{G0j}) in the time domain gives
\begin{eqnarray}
\lefteqn{ {\bf G}^R_{0,j}(t,t') = }
\nonumber \\
&& \int {dt_1} \int {dt_2} {\bf g}^R_{0,0}(t,t_1)
\left[
\left(
{\bf t}_{0,1} + i \hbar {\bf S}_{0,1} \frac{\partial}{\partial t_2}
\right)
\delta(t_1 - t_2)
\right]
{\bf G}^R_{1,j}(t_2,t') .
\label{GRt2}
\end{eqnarray}
Looking at the effective potential in the brackets, it is
still local in time. Or, generalizing this to the Keldysh
contour, $t_1$ and $t_2$ could never be on opposite
branches of the contour.
Therefore, it appears reasonable that no new self-energy
terms are introduced by the effective potential
$\tilde{\bf t}$ and that
by replacing ${\bf t}$ with $\tilde{\bf t}$,
$G^<$ and $\Sigma^<$ can be derived from the
usual Dyson equations for $G^<$
as described in detail in ref. \cite{LakeNemoTheory}.

Briefly, for completeness, we write down the Dyson equations for
$G^<$ including only the coupling to the left contact since
the equations to include coupling to the right contact are identical in
form and can be obtained by replacing subscript $1$ with $N$
and subscript $0$ with $N+1$.
For any atoms $i,j \in \{1, \ldots , N\}$ in the device,
\begin{equation}
{\bf G}_{i,j}^< =
\bgl_{i,j}
+
\bgr_{i,1} \btt_{1,0} \bGl_{0,j}
+
\bgl_{i,1} \btt_{1,0} \bGA_{0,j} .
\label{G<_ij}
\end{equation}
The Dyson equations for the exact $\bGl_{0,j}$ and $\bGA_{0,j}$ which
cross the device-contact boundary are
\begin{equation}
{\bf G}_{0,j}^< =
\bgr_{0,0} \btt_{0,1} \bGl_{1,j}
+
\bgl_{0,0} \btt_{0,1} \bGA_{1,j} .
\label{G<_0j}
\end{equation}
and
\begin{equation}
\bGA_{0,j} = \bga_{0,0} \btt_{0,1} \bGA_{1,j}
\label{GA_0j}
\end{equation}
Substituting Eqs. (\ref{GA_0j}) and (\ref{G<_0j}) back into Eq. (\ref{G<_ij}),
gives the exact expression for $\bGl$ of the device.
\begin{equation}
{\bf G}_{i,j}^< =
\bgl_{i,j}
+
\bgr_{i,1}
\underbrace{\btt_{1,0} \bgr_{0,0} \btt_{0,1}}_{{\bf \Sigma}^{RB}_{1,1}}
\bGl_{1,j}
+
\bgr_{i,1}
\underbrace{\btt_{1,0} \bgl_{0,0} \btt_{0,1}}_{{\bf \Sigma}^{<B}_{1,1}}
\bGA_{1,j}
+
\bgl_{i,1}
\underbrace{\btt_{1,0} \bga_{0,0} \btt_{0,1}}_{{\bf \Sigma}^{AB}_{1,1}}
\bGA_{1,j}
\label{G<_ij_all}
\end{equation}
Multiplying Eq. (\ref{G<_ij_all}) by the $N \times N$ block matrix
of the isolated device,
$[ E{\bf S}_0 - {\bf H}_0 ] = \left[ {\bf g}^R \right]^{-1}$,
and using $[ E{\bf S}_0 - {\bf H}_0 ] \bgl = 0$,
we obtain
the equation of motion for the contravariant correlation
function $G^<$ in the device domain
$i,j \in \{1, \ldots, N\}$.
\begin{equation}
\left[ E{\bf S}_0 - {\bf H}_0 -  {\bf \Sigma}^{RB} \right]
{\bf G}^< =
{\bf \Sigma}^{<B} {\bf G}^A
\label{eomG<}
\end{equation}
${\bf \Sigma}^{<B}$ is obtained from ${\bf \Sigma}^{RB}$, Eqs.
(\ref{Sigma_RL}) and (\ref{Sigma_RR}),
by replacing
${\bf g}^R$ with ${\bf g}^<$, i.e.
\begin{equation}
{\bf \Sigma}^{<B}_{1,1} = \left( {\bf t}_{1,0} - E{\bf S}_{1,0} \right)
{\bf g}^<_{0,0}
 \left( {\bf t}_{0,1} - E{\bf S}_{0,1} \right)
\label{Sigma_<L}
\end{equation}
\begin{equation}
{\bf \Sigma}^{<B}_{N,N} = \left( {\bf t}_{N,N+1} - E{\bf S}_{N,N+1} \right)
{\bf g}^<_{N+1,N+1}
 \left( {\bf t}_{N+1,N} - E{\bf S}_{N+1,N} \right)
\label{Sigma_<R}
\end{equation}

The one question left is what is
${\bf g}_{0,0}^<(E)$
where for any 2 orbitals $i,j$ associated with
atom $0$,
$g^<_{i_0,j_0}(E) = i
\langle b^{\dagger}_{j_0} b_{i_0}(E) \rangle$?
Previously, in an orthonormal basis, we say that
$b^{\dagger}_{j_0}$ creates the state $\ket{\alpha_{j_0}}$
localized in the left contact which is in equilibrium
by definition.
Therefore, ${\bf g}^<_{0,0}(E)$ is simply the
spectral function times the Fermi factor,
\begin{equation}
{\bf g}^<_{0,0}(E) = i f(E-\mu_L) {\bf a}_{0,0}(E)
\label{g<00}
\end{equation}
where the spectral function
${\bf a} = i\left[ {\bf g}^R - \left( {\bf g}^R \right)^{\dagger} \right]$,
and $f(E-\mu_L)$ is the Fermi factor for an electrochemical
potential of the left contact $\mu_L$.
Now, however, $b^{\dagger}_{j_0}$ creates the state
\begin{equation}
\ket{\beta_{j_0}} = \sum_n
\sum_{i} \left[S^{-1}\right]_{j_0,i_n} \ket{\alpha_{i_n}}
\label{Beta0}
\end{equation}
which is extended.
However, $g^<_{0,0}(E)$ is the noninteracting correlation function
for which there is no overlap between the contact and device states,
specifically, in our example, $\bS_{0,1} = \bS_{1,0} = 0$.
Therefore, $\bS^{-1}$ is block diagonal with no matrix elements
coupling the contacts to the device
and the sum in (\ref{Beta0}) is nonzero only for atoms $n$ in the
left contact, $n \in \{-\infty, \ldots, 0\}$.
Therefore, for the noninteracting ${\bf g}^<_{0,0}$,
$b^{\dagger}_{j_0}$ and $b_{i_0}$ only create and annihilate states
in the left contact which is in equilibrium so that
(\ref{g<00}) still holds.
The equations for the correlation function $G^<$ and Green function
$G^R$ are now formally identical
to those derived previously for orthogonal basis sets \cite{LakeNemoTheory}.

\subsection{Current}
\subsubsection{Standard Expression}

The derivation of the current operator generally starts with
an expression for the local electron density which is then placed into
the continuity equation to give an expression for the local divergence
of the current \cite{Caroli_I}.
Instead of a local continuity equation,
we will write a continuity equation for the total number of electrons
contained in the orbitals that are elements of the ``device.''
Note that we are defining the ``device'' as a set of orbitals
rather than a region of space.
The total electron number is
\begin{eqnarray}
N_D&=&-i \hbar \sum_{a,b = 1}^N
\sum_{i_a,j_b} G^<_{i_a,j_b}(t,t) \int d^3r \phi_{j_b}^*(\rv) \phi_{i_a}(\rv)
\nonumber \\
&=&
-i \hbar \: {\rm tr} \left\{ {\bf S}_0 {\bf G}^<(t,t) \right\}
\label{N_D}
\end{eqnarray}
where $a$ and $b$ index the atoms in the ``device,''
$i_a$ indexes an orbital of atom $a$, and
the trace is over all states in the ``device.''
We now take the time derivative of $N_D$ as
\begin{equation}
\frac{\partial}{\partial t} N_D =
-i \hbar \: {\rm tr} \:
\left\{
{\bf S}_0
\lim_{t' \rightarrow t}
\left[
\frac{\partial}{\partial t} + \frac{\partial}{\partial t'}
\right]
{\bf G}^<(t,t')
\right\}
\end{equation}
The equation of motion for ${\bf G}^<(t,t')$ in the device is
\begin{eqnarray}
\lefteqn{
i\hbar \frac{\partial}{\partial t} {\bf S}_0{\bf G}^<(t,t') -
{\bf H}_0 {\bf G}^<(t,t') = }
\nonumber \\
&&
\;\;\;\;\;\;
\;\;\;\;\;\;
\;\;\;\;\;\;
\int dt_1
\left[
{\bf \Sigma}^{RB} (t,t_1) {\bf G}^<(t_1,t')
+
{\bf \Sigma}^{<B}(t,t_1) {\bf G}^A(t_1,t')
\right]
\label{eomG<t}
\end{eqnarray}
and the conjugate equation is \cite{MahanArticle}
\begin{eqnarray}
\lefteqn{
-i \hbar \frac{\partial}{\partial t'} {\bf G}^<(t,t') {\bf S}_0 -
{\bf G}^<(t,t') {\bf H}_0 = }
\nonumber \\
&&
\;\;\;\;\;\;
\;\;\;\;\;\;
\;\;\;\;\;\;
\int dt_1
\left[
{\bf G}^<(t,t_1) {\bf \Sigma}^{AB} (t_1,t')
+
{\bf G}^R(t,t_1) {\bf \Sigma}^{<B}(t_1,t')
\right]
\label{eomG<t'}
\end{eqnarray}
Subtracting Eq. (\ref{eomG<t}) from Eq. (\ref{eomG<t'}),
taking the limit $t' \rightarrow t$, and the tracing over all
device states, we have
\begin{eqnarray}
\lefteqn{
\frac{\partial N_D}{\partial t}
+
{\rm tr}
\left\{
{\bf H}_0 {\bf G}^<(t,t) -
{\bf G}^<(t,t) {\bf H}_0
\right\}
- }
\nonumber \\
&&
\;\;\;\;\;\;
\;\;\;\;\;\;
\;\;\;\;\;\;
\;\;\;\;\;\;
\int dt_1
{\rm tr}
\{
{\bf G}^<(t,t_1) {\bf \Sigma}^{AB} (t_1,t)
+
{\bf G}^R(t,t_1) {\bf \Sigma}^{<B}(t_1,t)
\nonumber \\
&&
\;\;\;\;\;\;\;\;
\;\;\;\;\;\;\;\;
\;\;\;\;\;\;
\;\;\;\;\;\;
\;\;\;\;\;\;
\;\;\;\;\;\;
-
{\bf \Sigma}^{RB} (t,t_1) {\bf G}^<(t_1,t)
-
{\bf \Sigma}^{<B}(t,t_1) {\bf G}^A(t_1,t)
\} = 0 .
\label{continuity_t}
\end{eqnarray}
Eq. (\ref{continuity_t}) is the continuity equation for the total
electron number within the orbitals that define the ``device.''
In steady-state, $\frac{\partial N_D}{\partial t} = 0$.
The second term is also 0 due to the
cyclic invariance of the trace.
The last term must also be zero, but we can break it up into
contributions from the left and right contacts.
These contributions will be equal in magnitude and opposite in sign.
For our linear chain nearest neighbor example,
the self-energies resulting from coupling of the device
to the left contact are non-zero only for the orbitals
of atom 1.
Similarly, the self-energies resulting from coupling
of the device to the right contact are only non-zero
for the orbitals of atom N.

Fourier transforming the last term of Eq. (\ref{continuity_t}),
the particle current flowing out of the left contact into the device is
\begin{eqnarray}
\lefteqn{
J = \int \frac{dE}{2 \pi \hbar}
{\rm tr} \: \{ {\bf G}_{1,1}^<(E) {\bf \Sigma}_{1,1}^{AB}(E)
+
{\bf G}_{1,1}^R(E) {\bf \Sigma}_{1,1}^{<B}(E)
}
\nonumber \\
&&
\;\;\;\;\;\;\;\;\;\;\;\;\;\; \;\;\;\;\;\;\;\;\;\;\;\;\;\;
-
{\bf \Sigma}_{1,1}^{RB}(E) {\bf G}_{1,1}^<(E)
-
{\bf \Sigma}_{1,1}^{<B}(E) {\bf G}_{1,1}^A(E) ] \}
\label{JLE}
\end{eqnarray}
where the trace is over the orbitals of atom 1.
We regroup terms using the cyclic invariance of the trace to obtain
\begin{eqnarray}
J&=& \int \frac{dE}{2 \pi \hbar} {\rm tr} \:
\left\{
{\bf \Sigma}_{1,1}^{<B} \left( {\bf G}_{1,1}^R - {\bf G}_{1,1}^A \right)
-
{\bf G}_{1,1}^< \left( {\bf \Sigma}_{1,1}^{RB} - {\bf \Sigma}_{1,1}^{AB} \right)
\right\}
\nonumber \\
&=&
\int \frac{dE}{2 \pi \hbar}
{\rm tr} \:
\left\{
{\bf \Gamma}^B_{1,1} \left[ f_L {\bf A}_{1,1} + i {\bf G}^<_{1,1} \right] .
\right\}
\label{JL_MeirWingreen}
\end{eqnarray}
This current expression was first written down by
Meir and Wingreen \cite{Meir_Wingreen}.
In Eq. (\ref{JL_MeirWingreen}),
$f_L$ is the Fermi factor of the left contact,
${\bf \Gamma}^B_{1,1} = i\left( {\bf \Sigma}_{1,1}^{RB} - {\bf \Sigma}_{1,1}^{AB} \right)$,
and
${\bf A}_{1,1} = i\left( {\bf G}_{1,1}^{R} - {\bf G}_{1,1}^{A} \right)$
where, for example, ${\bf G}_{1,1}^{R}$ is the sub-block matrix
of ${\bf G}^R$ consisting
of orbitals of atom 1.

The integrand of Eq. (\ref{JL_MeirWingreen}) satisfies the
following equality \cite{LakeNemoTheory,Meir_Wingreen}
\begin{equation}
{\bf \Gamma}^B_{1,1} \left[
f_L {\bf A}_{1,1} + i {\bf G}^<_{1,1}
\right]
=
\btt_{0,1} \bGl_{1,0} - \btt_{1,0} \bGl_{0,1}
\label{T_G<10}
\end{equation}
This is derived by inserting the Dyson equations for
$\bGl_{0,1}$ and $\bGl_{1,0}$
\begin{eqnarray}
\bGl_{0,1}&=&
\bgr_{0,0} \btt_{0,1} \bGl_{1,1} + \bgl_{0,0} \btt_{0,1} \bGA_{1,1}
\label{G<01}
\\
\bGl_{1,0}&=&
\bGR_{1,1} \btt_{1,0} \bgl_{0,0} + \bGl_{1,1} \btt_{1,0} \bga_{0,0}
\label{G<10}
\end{eqnarray}
into the right hand side of Eq. (\ref{T_G<10}).
Therefore, the general expression for the current
Eq. (\ref{JL_MeirWingreen}) is also given by
\begin{equation}
J
=
\int \frac{dE}{2 \pi \hbar} {\rm tr} \:
\left\{
\btt_{0,1} \bGl_{1,0} - \btt_{1,0} \bGl_{0,1}
\right\}
\label{J_G<10}
\end{equation}
Since there is nothing special about atoms 0 and 1,
we can say that Eq. (\ref{JL_MeirWingreen}) implies that
the current flowing between any 2 atoms $n$ and $n+1$ in the chain is
\begin{equation}
J = \int \frac{dE}{2 \pi \hbar}
{\rm tr}
\left\{
\btt_{n,n+1} \bGl_{n+1,n} - \btt_{n+1,n} \bGl_{n,n+1}
\right\}
\label{J_general}
\end{equation}
or explicitly writing out the $\btt$ terms,
\begin{equation}
J = \int \frac{dE}{2 \pi \hbar}
{\rm tr}
\left\{
\left(\bt_{n,n+1} -E\bS_{n,n+1}\right) \bGl_{n+1,n} -
\left(\bt_{n+1,n} -E\bS_{n+1,n}\right) \bGl_{n,n+1}
\right\}
\label{J_general_explicit}
\end{equation}

Eqs. (\ref{J_G<10}), (\ref{J_general}) and (\ref{J_general_explicit}) are
identical to those obtained from empirical tight binding
models \cite{Caroli_I,LakeNemoTheory}.

The matrix elements $(\bt - E\bS)/\hbar$ can be thought of as a transition rate
coupling orbitals from atoms $n$ and $n \pm 1$.
It is not immediately clear that $E\bS$ should enter into the dynamics.
We will show below that these are the correct matrix elements
which result in unitary transmission probabilities for a band
of a periodic system.

For coherent transport, ${\bf G}^< = {\bf G}^R {\bf \Sigma}^{<B} {\bf G}^A$
and
${\bf A} = {\bf G}^R  {\bf \Gamma}^B {\bf G}^A$ or
writing out these expressions explicitly,
\begin{equation}
{\bf G}^<_{1,1}
=
\bGR_{1,1} \left( i f_L  \BG^B_{1,1} \right) \bGA_{1,1}
+
\bGR_{1,N} \left( i f_R  \BG^B_{N,N} \right) \bGA_{N,1}
\label{G<coherent}
\end{equation}
and
\begin{equation}
{\bf A}_{1,1}
=
\bGR_{1,1} \BG^B_{1,1}  \bGA_{1,1}
+
\bGR_{1,N} \BG^B_{N,N} \bGA_{N,1}.
\label{A_coherent}
\end{equation}
Inserting the above two equations into Eq. (\ref{JL_MeirWingreen})
results in the
usual form for the
coherent current \cite{Meir_Wingreen,LakeNemoTheory,FisherLee}.
\begin{equation}
J =
\int \frac{dE}{2 \pi \hbar}
{\rm tr}
\left\{
{\bf \Gamma}_{1,1}  {\bf G}^R_{1,N} {\bf \Gamma}_{N,N} {\bf G}^A_{N,1}
\right\}
\left( f_L - f_R \right)
\label{FisherLee}
\end{equation}
This equation was first derived by Caroli et al. \cite{Caroli_I}
(see their Eq. (42)) and then later by Fisher and Lee
\cite{FisherLee} and then may others \cite{Meir_Wingreen}.

Once one has derived an expression for the transmission coefficient,
it is always wise to check that for a periodic, uniform system, the
transmission is 1.0 independent of energy within a propogating band.
We will now perform this check of Eq. (\ref{FisherLee}) for a linear chain
of atoms with one orbital per atom.
The Hamiltonian matrix elements are
$\bra{\alp_n} H \ket{\alp_n} = \epsilon$ and
$\bra{\alp_n} H \ket{\alp_{n \pm 1}} = t$
where $n$ is the atom index.
The overlaps are $\braket{\alp_n}{\alp_{n \pm 1}} = s$
and $\braket{\alp_n}{\alp_n} = 1$.
Applying the Hamiltonian to the {\em ansatz}
$\ket{\psi} = \sum_n \ket{\alp_n} z^n$ results in
\begin{equation}
E = \left( t-Es \right) z^{-1} + \ep + \left( t-Es\right) z
\label{E_z}
\end{equation}
Substituting the phase factor $z = e^{ik}$ into
(\ref{E_z}) results in the $E-k$ dispersion relation
\begin{equation}
E - \ep + 2 \left( t - Es \right) \cos (k) = 0 .
\label{dispersion}
\end{equation}
Since we are considering a uniform infinite chain of atoms, our
device will consist of 1 atom, $n=1$.
The surface Green function $g^R_s = g^R_{0,0} = g^R_{2,2}$
of both the left and right leads
is given by the recursion relation
\begin{equation}
g^R_s = \left[ E - \ep - \left( t - Es \right) g^R_s \left( t - Es \right) \right]^{-1}
\label{gRs}
\end{equation}
Eq. (\ref{gRs}) simply states that the surface Green function
of the semi-infinite chain ending at site 0 is the same as the
surface Green function
of the semi-infinite chain ending at site -1.
Eq. (\ref{E_z}) is a quadratic equation for $z$ and Eq. (\ref{gRs})
is a quadratic equation for $g^R_s$.
Writing out the solution for both equations, one finds that
\begin{equation}
g^R_s = \frac{z}{t-Es} = \frac{e^{ik}}{t-Es} = \frac{\cos(k) + i\sin(k)}{t-Es}
\label{gRs_z}
\end{equation}
The boundary self-energy for both the left and right contact is
\begin{equation}
\Sigma^{RB} = \left(t - Es \right) g^R_s \left(t - Es \right)
\label{Sigma_RB_scalar}
\end{equation}
and
\begin{equation}
\Gamma^B = \Gamma^L = \Gamma^R = -2 {\rm Im} \: \Sigma^{RB} = -2 \left(t - Es \right) \sin(k)
\label{GammaB_scalar}
\end{equation}
The superscripts $L$ and $R$ indicate that the quantity is the result of
coupling to the left or right contact, respectively.
The exact Green function on site 1 is
\begin{eqnarray}
G^R_{1,1}&=&\left[E - \ep - 2 \left(t - Es \right) g^R_s \left(t - Es \right) \right]^{-1}
\nonumber \\
&=&
\left[E - \ep - 2 \left(t - Es \right)\left( \cos(k) + i\sin(k) \right) \right]^{-1}
\nonumber \\
&=&
\left[- 2 i \left(t - Es \right) \sin(k) \right]^{-1}
\nonumber \\
&=&
\frac{1}{i \frac{1}{2} \left(\Gamma^L + \Gamma^R \right) }
= \frac{1}{i \Gamma^B} .
\label{GR_site_1}
\end{eqnarray}
The factor of 2 comes from the
two identical self energies from the left and right
leads.
In the second to last line, the real part is zero
from the dispersion relation Eq. (\ref{dispersion}).
The transmission coefficient from Eq. (\ref{FisherLee})
is $T = \Gamma^L_{1,1} G^R_{1,1} \Gamma^R_{1,1} G^{R^*}_{1,1}$
which, for scalar quantities, is
\begin{equation}
T = {\Gamma^{B}}^2 \left| G^R_{1,1} \right|^2
\label{T_scalar}
\end{equation}
where $\Gamma^B = \Gamma^L = \Gamma^R$.
Substituting Eq. (\ref{GR_site_1}) into
Eq. (\ref{T_scalar}) results in
a unitary transmission probability.
\begin{equation}
T = \frac{{\Gamma^{B}}^2 }{{\Gamma^{B}}^2 } \:=\: 1
\label{T_scalar_final}
\end{equation}
This demonstrates that the matrix elements $\bt - E\bS$ are the
correct ones to use in the current equations.

With these expressions for the current Eqs. (\ref{JL_MeirWingreen}),
(\ref{J_G<10}), (\ref{J_general_explicit}), and
(\ref{FisherLee}), we have re-derived
NEGF theory for a non-orthogonal basis that is formally
identical to that for an orthogonal basis.
The equations for the correlation function $G^<$, the Green function
$G^R$, and the current $J$, are now formally identical
and can be obtained by simply replacing the off
diagonal Hamiltonian matrix elements ${\bf t}$ with
${\bf t}-E{\bf S}$.
All of the efficient numerical algorithms developed previously,
such as the
current expressions
and
recursive Green function algorithms,
can be applied \cite{LakeNemoTheory,Svizhenko_JAP02,Dresden02}.

\subsubsection{Direct Evaluation of Current Operator}
In deriving the standard current equation (\ref{JL_MeirWingreen}), we
started by calculating the current flowing out a specific set of
orbitals rather than a region of space. Since our model has a
specific basis, it would appear that
we do not need to make this approximation,
and we can directly calculate the surface integral of the
current crossing a plane.
However, we will
demonstrate by example that for an incomplete
basis
such a straightforward
calculation results in an
expression for the current which violates particle conservation
and does not reduce
to unitary transmission for coherent transport within a band.

For evaluating  the surface integral of the
current crossing a plane,
we choose the plane between the atoms 0 and 1, i.e. between the
left contact and the device.
The expectation value of the current crossing the plane is
\begin{eqnarray}
J = \int d {\bf s} \cdot {\bf J}(\rv)
&=&
\frac{-i \hbar}{2 m} \int d {\bf s} \cdot
\langle
\psi^{\dagger}(\rv) \nabla \psi(\rv)
-
\left[ \nabla \psi^{\dagger}(\rv) \right] \psi(\rv)
\rangle
\label{J_field_ops}
\\
&=&
\frac{-\hbar^2}{2 m} \int d {\bf s} \cdot
\lim_{\rv' \rightarrow \rv}
\left( \nabla - \nabla^{\prime} \right)
G^<(\rv,t; \rv',t)
\label{J_x}
\\
&=&
\frac{-i \hbar}{2 m}
\sum_{i, j}
\langle
b_{j}^{\dagger} b_{i}
\rangle
\int d {\bf s} \cdot
\left[
\phi_{j}^*(\rv_2) \nabla \phi_{i}(\rv)
-
\phi_{i}(\rv) \nabla \phi_{j}^*(\rv)
\right]
\nonumber \\
&=&
\frac{-\hbar^2}{2 m}
\sum_{i, j}
G^<_{i,j}(t,t)
\int d {\bf s} \cdot
\left[
\phi_{j}^*(\rv) \nabla \phi_{i}(\rv)
-
\phi_{i}(\rv) \nabla \phi_{j}^*(\rv)
\right]
\label{Jdots}
\end{eqnarray}
Eq. (\ref{Jdots}) is the one that we will use. We write
it in the form of Eq. (\ref{J_x}) to
show that for a complete basis, the value of the surface integral
is independent of position along the wire.

We take the plane of integration to be the $xy$ plane. Fourier transforming
into the energy domain and taking $\partial / \partial z$ of
Eq. (\ref{J_x}) results in
\begin{equation}
\frac{\partial}{\partial z} J =
\frac{-\hbar^2}{2 m} \int \frac{dE}{2 \pi \hbar}
\int dx \int dy
\lim_{\rv' \rightarrow \rv}
\left( \frac{\partial^2}{\partial z^2} - \frac{\partial^2}{\partial {z'}^2} \right)
G^<(\rv, \rv';E)
\label{J_x_E}
\end{equation}
Using the equations of motion for $G^<(\rv, \rv';E)$ within the device
\begin{equation}
\left[ E + \frac{\hbar^2 \nabla^2}{2m} - V(\rv) \right]G^<(\rv, \rv';E) = 0
\end{equation}
\begin{equation}
\left[ E + \frac{\hbar^2 {\nabla'}^2}{2m} - V(\rv') \right]G^<(\rv, \rv';E) = 0
\end{equation}
Eq. (\ref{J_x_E}) becomes
\begin{equation}
\frac{\partial}{\partial z} J =
\frac{\hbar^2}{2 m}
\int \frac{dE}{2 \pi \hbar}
\int dx \int dy
\lim_{\rv' \rightarrow \rv}
\left(
\frac{\partial^2}{\partial x^2} + \frac{\partial^2}{\partial y^2}
- \frac{\partial^2}{\partial {x'}^2}
- \frac{\partial^2}{\partial {y'}^2}
\right)
G^<(\rv, \rv';E)
\label{J_xy_E}
\end{equation}
Integrating by parts using
\begin{equation}
\frac{\partial}{\partial x} G^<(\rv, \rv ;E) =
\lim_{\rv' \rightarrow \rv}
\left(
\frac{\partial}{\partial x} + \frac{\partial}{\partial x'}
\right)
G^<(\rv, \rv' ;E)
\end{equation}
Eq. (\ref{J_xy_E}) is equal to zero.
This is perhaps easier to see by re-writing Eq. (\ref{J_xy_E}) back in
the form of Eq. (\ref{J_field_ops}).

We now return to Eq. (\ref{Jdots}) which we write out explicitly
as
\begin{equation}
J =
\sum_{i, j}
G^<_{i,j}(t,t)
\int dx \int dy
\left( \frac{-\hbar^2}{2 m} \right)
\left[
\phi_{j}^*(\rv) \frac{\partial}{\partial z} \phi_{i}(\rv)
-
\phi_{i}(\rv) \frac{\partial}{\partial z} \phi_{j}^*(\rv)
\right]
\label{J_dxdy}
\end{equation}
We evaluate (\ref{J_dxdy}) at $z$ between sites $0$ and $1$.
The only nonzero contributions come from the orbitals that
cut the surface between atoms 0 and 1.
In our nearest neighbor example, only the orbitals of atoms
0 and 1 contribute to the integral.
The integral in (\ref{J_dxdy}) contains 4 terms,
2 intra-atomic terms and 2 inter-atomic
terms. We will first consider the inter-atomic terms since these
2 terms will result in an expression formally identical
to Eq. (\ref{J_G<10}) but with different matrix elements.

The inter-atomic terms are
\begin{equation}
J_{t} =
\sum_{i_0, j_1}
\left[ M_{i_0,j_1} - M^*_{j_1, i_0} \right]
G^<_{j_1, i_0}(t,t)
+
\left[ M_{j_1,i_0} - M^*_{i_0,j_1} \right]
G^<_{i_0,j_1}(t,t)
\label{Jt}
\end{equation}
where $M_{i,j}$
is the matrix element
$\int dx \int dy \left( \frac{-\hbar^2}{2 m} \right) \phi_{i}^*(\rv) \frac{\partial}{\partial z} \phi_{j}(\rv)$.

We define the current
matrix element as
\begin{equation}
J_{i_0,j_1} = \left[ M_{i_0,j_1} - M^*_{j_1, i_0} \right]
\label{CurrentMatrixElt}
\end{equation}
and Eq. (\ref{Jt}) becomes
\begin{eqnarray}
J_{t} &=&
\sum_{i_0, j_1}
\left[
J_{i_0,j_1}  G^<_{j_1,i_0}(t,t)
+
J_{j_1,i_0} G^<_{i_0,j_1} (t,t)
\right]
\nonumber \\
&=&
{\rm tr} \:
\left\{
\bJ_{0,1} \bGl_{1,0}(t,t)
+
{\bf J}_{1,0} {\bf G}^<_{0,1}(t,t)
\right\}
\nonumber \\
&=&
\int \frac{dE}{2 \pi \hbar}
{\rm tr} \:
\left\{
\bJ_{0,1} \bGl_{1,0}(E)
-
{\bf J}^{\dagger}_{0,1} {\bf G}^<_{0,1}(E)
\right\}
\label{Jt2}
\end{eqnarray}
where in the last line we used $\bJ_{0,1}^{\dagger} = - \bJ_{1,0}$.

Eq. (\ref{Jt2}) is a general expression for the current.
We can move the plane to cut between any
two atoms of the chain, so that we have an expression
for calculating the current at any point in the device.
For example, cutting between atoms $n$ and $n+1$,
the current would be
\begin{equation}
J_t =
\int \frac{dE}{2 \pi \hbar}
{\rm tr} \:
\left\{
\bJ_{n,n+1} \bGl_{n+1,n}(E)
-
{\bf J}^{\dagger}_{n,n+1} {\bf G}^<_{n,n+1}(E)
\right\} .
\label{Jtn}
\end{equation}

From now on, we will work in the energy domain and
save on notation by defining $J_t(E)$ as
$J_t(E) = {\rm tr} \: \left\{ \bJ_{0,1} \bGl_{1,0}(E) - {\bf J}^{\dagger}_{0,1} {\bf G}^<_{0,1}(E) \right\}$.
We will also
use the matrix notation of the last line of Eq.
(\ref{Jt2}).
The bold quantities indicate matrices
the size of the localized orbital basis per atom and
the subscripts indicate from which atom the orbitals come.
The trace is over the orbitals equivalent to the
$\sum_{i_0, j_1}$ in Eq. (\ref{Jt2}).

To bring Eq. (\ref{Jt2}) into a form comparable to Eq. (\ref{JL_MeirWingreen})
or Eq. (\ref{FisherLee}),
we begin by
substituting the Dyson Equations for $\bGl_{0,1}$ and
$\bGl_{1,0}$
Eqs. (\ref{G<01}) and (\ref{G<10}) into Eq. (\ref{Jt2})
to obtain
\begin{equation}
J_t(E) = {\rm tr}
\left\{ \bJ_{0,1} \bGR_{1,1} \btt_{1,0} \bgl_{0,0}
+
\bJ_{0,1} \bGl_{1,1} \btt_{1,0} \bga_{0,0}
-
\bJ^{\dagger}_{0,1} \bgr_{0,0} \btt_{0,1} \bGl_{1,1}
-
\bJ^{\dagger}_{0,1} \bgl_{0,0} \btt_{0,1} \bGA_{1,1}
\right\} .
\label{JtG<exp}
\end{equation}
Eq. (\ref{JtG<exp}) is the equivalent of Eq. (\ref{JL_MeirWingreen}).
If the $\bJ$s were replaced by $\btt$s, then
Eq. (\ref{JtG<exp}) would be identical to Eq. (\ref{JL_MeirWingreen}).
Grouping the terms of Eq. (\ref{JtG<exp}) with $\bGl_{1,1}$
together and using the cyclic invariance of the trace, Eq. (\ref{JtG<exp})
becomes
\begin{eqnarray}
J_t(E)&=& {\rm tr} \:
\left\{
i\: \left[ \left( \bJ^{\dagger}_{0,1} \bgr_{0,0} \btt_{0,1} \right)
-
\left( \bJ^{\dagger}_{0,1} \bgr_{0,0} \btt_{0,1} \right)^{\dagger}
\right]
\: i \bGl_{1,1}
\right\}
\nonumber \\
&&
\;\;\;\;\;\;\;\;\;\;\;\;\;\;\;\;\;\;
+  {\rm tr} \:
\left\{ \btt_{1,0} \bgl_{0,0} \bJ_{0,1}  \bGR_{1,1}
-
\bJ^{\dagger}_{0,1} \bgl_{0,0} \btt_{1,0}  \bGA_{1,1}
\right \}
\nonumber \\
&=&
{\rm tr} \: \mbox{\boldmath $\gamma$}^J_{1,1} \: i \bGl_{1,1}
+  {\rm tr} \:
\left\{ \btt_{1,0} \bgl_{0,0} \bJ_{0,1}  \bGR_{1,1}
-
\bJ^{\dagger}_{0,1} \bgl_{0,0} \btt_{1,0}  \bGA_{1,1}
\right\}
\label{JtG<exp3}
\end{eqnarray}
where we used $\btt_{0,1}^{\dagger} = \btt_{1,0}$,
$[ \bgr_{0,0}]^{\dagger} = \bga_{0,0}$,
and defined
\begin{equation}
\mbox{\boldmath $\gamma$}^J_{1,1} \doteq
i\: \left[ \left( \bJ^{\dagger}_{0,1} \bgr_{0,0} \btt_{0,1} \right)
-
\left( \bJ^{\dagger}_{0,1} \bgr_{0,0} \btt_{0,1} \right)^{\dagger}
\right].
\label{gammaJdef}
\end{equation}
Eq. (\ref{JtG<exp3}) is exact, valid with or without incoherent
processes in the device.

The 2 intra-atomic terms from Eq. (\ref{J_dxdy}) result in a current
term
\begin{equation}
J_a = \int \frac{dE}{2 \pi \hbar}
{\rm tr} \:
\left\{
{\bf J}_{0,0} {\bf G}^<_{0,0}(E)
+
\bJ_{1,1} \bGl_{1,1}(E)
\right\}
\label{J_intra_atomic}
\end{equation}
In equilibrium when $\bGl = if\bA$, the intra-atomic terms are individually 0
since for our real orbitals,
$\bA(t,t) = {\bf S}^{-1}$ is symmetric and $\bJ$ is antisymmetric.
${\bf G}^<_{0,0}(E)$ is the exact ${\bf G}^<$ in the left contact,
which is, since the contact is by definition in equilibrium,
${\bf G}^<_{0,0}(E) = i f_L \bA_{0,0}(E)$ \cite{continuum_approx}.
Therefore, the first term of Eq. (\ref{J_intra_atomic}) is always zero
for a basis of real orbitals.

For coherent transport, we can bring the sum of Eqs. (\ref{JtG<exp3})
and Eq. (\ref{J_intra_atomic}) into the Fisher-Lee form for the
transmission coefficient.
Substituting \cite{continuum_approx}
\begin{equation}
\bGl_{1,1} = i f_L \bGR_{1,1} \BG_{1,1}^B \bGA_{1,1}
+
i f_R \bGR_{1,N} \BG_{N,N}^B \bGA_{N,1}
\label{G<11}
\end{equation}
into Eq. (\ref{JtG<exp3}) gives
\begin{eqnarray}
\lefteqn{
J_t(E)= - {\rm tr} \:
\{\mbox{\boldmath $\gamma$}^J_{1,1} \bGR_{1,N} \BG_{N,N}^B \bGA_{N,1} \} f_R
}
\nonumber \\
&&
\;\;\;\;\;\;\;\;\;\;\;\;\;\;\;\;\;\;
+  {\rm tr} \:
\left\{
- \mbox{\boldmath $\gamma$}^J_{1,1}  \bGR_{1,1} \BG_{1,1}^B \bGA_{1,1}
+
i
\left[
\btt_{1,0} \ba_{0,0}  \bJ_{0,1} \bGR_{1,1}
-
\bJ^{\dagger}_{0,1} \ba_{0,0} \btt_{0,1} \bGA_{1,1}
\right]
\right\} f_L .
\label{Jt_balance_trick}
\end{eqnarray}
where in the second line we used the relation
$\bgl_{0,0} = i f_L \ba_{0,0}$.
Substituting Eq. (\ref{G<11}) into the intra-atomic current
term Eq. (\ref{J_intra_atomic}) gives \cite{continuum_approx}
\begin{equation}
J_a = \int \frac{dE}{2 \pi \hbar}
{\rm tr} \:
\left\{
i f_L \bJ_{1,1} \bGR_{1,1} \BG_{1,1}^B \bGA_{1,1}
+
i f_R \bJ_{1,1} \bGR_{1,N} \BG_{N,N}^B \bGA_{N,1}
\right\}
\label{J_intra_atomic_coherent}
\end{equation}
The total coherent current is the sum of the intra-atomic
Eq. (\ref{J_intra_atomic_coherent})
and inter-atomic Eq. (\ref{Jt_balance_trick}) contributions.

In equilibrium, when $f_L$ = $f_R$, the current must be 0.
Furthermore, we can, in principle change $f_L$ and $f_R$
independently of the transmission coefficient.
Therefore, the term proportional to $f_L$ must be equal
in magnitude and opposite in sign to the term
proportional to $f_R$. Both terms are alternative
expressions for the transmission coefficient.
The term proportional to $f_R$ is in the standard
Fisher-Lee form so that we can write
\begin{equation}
J =  \int \frac{dE}{2 \pi \hbar}
{\rm tr} \:
\left\{
\left(
\mbox{\boldmath $\gamma$}^J_{1,1} - i \bJ_{1,1}
\right)
\bGR_{1,N} \BG_{N,N}^B \bGA_{N,1}
\right\}
\left( f_L - f_R \right)
\label{Jexact_FisherLee}
\end{equation}
We note that the other term of Eq. (\ref{Jt_balance_trick})
is the more computationally efficient form equivalent to
Eq. (53) of ref. \cite{LakeNemoTheory}.
Also, $\mbox{\boldmath $\gamma$}^J_{1,1}$ is Hermitian
and the product $\bGR_{1,N} \BG_{N,N}^B \bGA_{N,1}$
is Hermitian. Therefore, the trace
${\rm tr} \: \left\{ \mbox{\boldmath $\gamma$}^J_{1,1} \bGR_{1,N} \BG_{N,N}^B \bGA_{N,1} \right\} $
is real.
Conversely, $\bJ_{1,1}$ is antisymmetric so that
$i \: {\rm tr} \: \left\{\bJ_{1,1} \bGR_{1,N} \BG_{N,N}^B \bGA_{N,1} \right\}$
is also real.
Eq. (\ref{Jexact_FisherLee}) is identical to the expression
that is universally
used \cite{Seminario_NEGF01,Taylor_ab_init_PRB01,Xue_JChemPhys01,Ghosh_Datta_PRB01,Brandbyge02,Xue_ChemPhys02}
Eq. (\ref{FisherLee}) with
the replacement of
${\bf \Gamma}_{1,1}$ with
$( \mbox{\boldmath $\gamma$}^J_{1,1} -  i \bJ_{1,1} )$.

With current expression (\ref{Jexact_FisherLee})
we investigate the consequence of an incomplete basis
by evaluating
the value of the transmission probability for a periodic, uniform system
as we did for the ``standard'' current expression Eq. (\ref{FisherLee}).
Using the same single-band model described immediately preceding
Eq. (\ref{E_z}), the intra-atomic term $J_{1,1}$ is 0.
The factor $\gamma^J$ is
\begin{equation}
\gamma^J = -2 {\rm Im} \: J_{0,1} g_s^R (t-Es) = -2 J_{0,1} \sin(k) .
\end{equation}
The transmission coefficient Eq. (\ref{Jexact_FisherLee}) is then
\begin{equation}
T = \frac{J_{0,1}}{t-Es}.
\label{T_exact_ssb}
\end{equation}

The transmission probability of Eq. (\ref{T_exact_ssb})
can only equal unity if $J_{0,1} = t-Es$.
Since $t-Es$ is energy dependent, this is clearly not possible.
Therefore,
this ``exact'' evaluation of the current
leads to qualitatively incorrect results
for an incomplete basis.
With an incomplete basis, one should
derive the current expression from
conservation laws that are consistent
with the Hamiltonian and basis such as
Eqs. (\ref{N_D}) - (\ref{continuity_t}).
We note the strong parallel here with the discussions in the literature
concerning the form of the momentum matrix element in
empirical tight binding theory in which the explicit
basis is not
known \cite{Ram-Mohan_p,Graf_Vogl_E_M_fields,Boykin_eiqr,Boykin_Vogl}.

\subsubsection{Equivalence of the Standard and Exact Current Expressions in a Complete Basis}

The argument for the equivalence of the two current expressions,
Eqs. (\ref{JL_MeirWingreen}) or (\ref{J_G<10}) and Eqs. (\ref{Jt2}) plus
(\ref{J_intra_atomic}) is
based on particle conservation by
writing the current as the time derivative of the
electron number in the left contact \cite{korotkov}.
For lack of better expressions we will refer to the current
in the ``standard'' expressions Eqs. (\ref{JL_MeirWingreen}) or (\ref{J_G<10})
as the ``tight-binding current'' ($J_{tb}$).
We will refer to the current derived from the direct evaluation of the
current operator Eq. (\ref{J_x}) as the ``real space current'' ($J_z$).

For the real space current, Eq. (\ref{J_x}) is derived from the
continuity equation, and it is equal to the negative time derivative of the electron
number in the left contact (for the surface chosen between atoms 0 and 1).
It is given by
\begin{eqnarray}
J_z&=&- \frac{\partial N^L_z}{\partial t} =
- \frac{\partial}{\partial t} \int dx \int dy \int_{-\infty}^{z_0} dz \:
\langle \psi^{\dagger}(\rv) \psi(\rv) \rangle
\nonumber \\
&=&
i \hbar
\frac{\partial}{\partial t}
\sum_{i,j}
G^<_{i,j}(t,t)
\int dx \int dy \int_{-\infty}^{z_0} dz \:
\phi_j^*(\rv) \phi_i(\rv)
\label{dNLzdt}
\end{eqnarray}
where $N^L_z$ is the electron number of the left contact
and $z_0$ lies between atoms 0 and 1.

The tight binding current
Eqs. (\ref{JL_MeirWingreen}) or (\ref{J_G<10}) were derived
by considering the time derivative of the electron number of the
orbitals in the ``device.'' It can also be derived by considering
the time derivative of the electron number of the orbitals
in the left lead.
Although
this seems physically obvious, it is not straightforward to show,
therefore below, we write down the main steps.

The electron number in the left contact is given by the sum over all
of the orbitals of the left contact,
\begin{eqnarray}
N^L_{tb} &=&-i \hbar \sum_{a,\:b = -\infty}^0 \;\:\:\:
\sum_{i_a,j_b} G^<_{i_a,j_b}(t,t) \int d^3r \phi_{j_b}^*(\rv) \phi_{i_a}(\rv)
\nonumber \\
&=&
-i \hbar \: {\rm tr} \left\{ {\bf S}_0 {\bf G}^<(t,t) \right\}
\label{N_L}
\end{eqnarray}
where the trace is over all orbitals of the left contact.
We need an equation for the exact $G^<$ of the left contact in terms of the
exact Green functions of the device.
Mimicing the steps used to derive $G^<$ of the device
Eqs. (\ref{G<_ij}) - (\ref{eomG<})
we write the Dyson equations for the left contact.
For any atoms $i,j \in \{- \infty, \ldots , 0\}$ in the left contact,
\begin{equation}
{\bf G}_{i,j}^< =
\bgl_{i,j}
+
\bgr_{i,0} \btt_{0,1} \bGl_{1,j}
+
\bgl_{i,0} \btt_{0,1} \bGA_{1,j} .
\label{G<L_ij}
\end{equation}
The Dyson equations for the exact $\bGl_{1,j}$ and $\bGA_{1,j}$ which
cross the device-contact boundary are
\begin{equation}
{\bf G}_{1,j}^< =
\bGR_{1,1} \btt_{1,0} \bgl_{0,j}
+
\bGl_{1,1} \btt_{1,0} \bga_{0,j} .
\label{G<_1j}
\end{equation}
and
\begin{equation}
\bGA_{1,j} = \bGA_{1,1} \btt_{1,0} \bga_{0,j}
\label{GA_1j}
\end{equation}
Substituting Eqs. (\ref{GA_1j}) and (\ref{G<_1j}) back into Eq. (\ref{G<L_ij}),
gives the exact expression for $\bGl$ of the left lead.
\begin{equation}
{\bf G}_{i,j}^< =
\bgl_{i,j}
+
\bgr_{i,0} \btt_{0,1} \bGR_{1,1} \btt_{1,0} \bgl_{0,j}
+
\bgr_{i,0} \btt_{0,1} \bGl_{1,1} \btt_{1,g} \bga_{0,j}
+
\bgl_{i,0} \btt_{0,1} \bGA_{1,1} \btt_{1,0} \bga_{0,j}
\label{Dyson_G<L}
\end{equation}
Multiplying (\ref{Dyson_G<L}) on the left with
${\bgr}^{-1} = \left[ E \bS_0 - {\bf H}_0 \right]$
results in
\begin{equation}
\sum_{k=-\infty}^0 \left[ E \bS_0 - {\bf H}_0 \right]_{i,k} \bGl_{k,j}
=
\delta_{i,0}
\underbrace{\btt_{0,1} \bGR_{1,1} \btt_{1,0}}_{{\bf \Sigma}^{RB}_{0,0}}
\bgl_{0,j}
+
\delta_{i,0}
\underbrace{\btt_{0,1} \bGl_{1,1} \btt_{1,0}}_{{\bf \Sigma}^{<B}_{0,0}} \bga_{0,j} .
\end{equation}
In the time domain
we have
\begin{equation}
i \hbar \frac{\partial}{\partial t} {\bf S}_0{\bf G}^<(t,t') -
{\bf H}_0 {\bf G}^<(t,t') =
\int dt_1
\left[
{\bf \Sigma}^{RB} (t,t_1) {\bf g}^<(t_1,t')
+
{\bf \Sigma}^{<B}(t,t_1) {\bf g}^A(t_1,t')
\right]
\label{eomG<t_L}
\end{equation}
and
\begin{equation}
-i \hbar \frac{\partial}{\partial t'} {\bf G}^<(t,t') {\bf S}_0-
{\bf G}^<(t,t')
{\bf H}_0
=
\int dt_1
\left[
{\bf g}^<(t,t_1)
{\bf \Sigma}^{AB} (t_1,t')
+
{\bf g}^R(t,t_1)
{\bf \Sigma}^{<B}(t_1,t')
\right]
\label{eomG<tp_L}
\end{equation}
Subtracting Eq. (\ref{eomG<t_L}) from Eq. (\ref{eomG<tp_L}),
taking the limit $t' \rightarrow t$, and the tracing over all
contact states, we have
\begin{eqnarray}
\lefteqn{
\frac{\partial N^L_{tb}}{\partial t}
+
{\rm tr}
\left\{
{\bf H}_0 {\bf G}^<(t,t) -
{\bf G}^<(t,t) {\bf H}_0
\right\}
- }
\nonumber \\
&&
\;\;\;\;\;\;
\;\;\;\;\;\;
\;\;\;\;\;\;
\;\;\;\;\;\;
\int dt_1
{\rm tr}
\{
{\bf g}^<(t,t_1) {\bf \Sigma}^{AB} (t_1,t)
+
{\bf g}^R(t,t_1) {\bf \Sigma}^{<B}(t_1,t)
\nonumber \\
&&
\;\;\;\;\;\;\;\;
\;\;\;\;\;\;\;\;
\;\;\;\;\;\;
\;\;\;\;\;\;
\;\;\;\;\;\;
\;\;\;\;\;\;
-
{\bf \Sigma}^{RB} (t,t_1) {\bf g}^<(t_1,t)
-
{\bf \Sigma}^{<B}(t,t_1) {\bf g}^A(t_1,t)
\} = 0 .
\label{continuity_L}
\end{eqnarray}
The second term in Eq. (\ref{continuity_L}) is zero due to the
cyclic invariance of the trace.
Writing out the last term we get that the current flowing out
of the left contact is
\begin{eqnarray}
\lefteqn{
J_{tb} = \frac{\partial N^L_{tb}}{\partial t}
=
\int \frac{dE}{2 \pi \hbar}
{\rm tr} \: \{ \btt_{0,1} \bGR_{1,1} \btt_{1,0} \bgl_{0,0}
+
\btt_{0,1} \bGl_{1,1} \btt_{1,0} \bga_{0,0}
}
\nonumber \\
&&
\;\;\;\;\;\;\;\;\;\;\;\;\;\;\;\;\;
\;\;\;\;\;\;\;\;\;\;\;\;\;\;\;\;\;
\;\;\;\;\;\;\;\;\;\;\;\;\;\;\;\;\;
-
\bgl_{0,0} \btt_{0,1} \bGA_{1,1} \btt_{1,0}
-
\bgr_{0,0} \btt_{0,1} \bGl_{1,1} \btt_{1,0} \}.
\label{JL1}
\end{eqnarray}
Cyclically permuting the terms under the trace, Eq. (\ref{JL1})
is identical to
Eq. (\ref{JLE}) derived by considering the time derivative of the
electron number of the device.

Now that we have shown that both currents are equal to the negative
time derivative of the total electron number of the left contact,
we take the long time average of the current.
\begin{equation}
\langle J_z \rangle = \frac{1}{2\tau} \int_{-\tau}^{\tau}{dt} J_z
=
-\left[ N^L_z(\tau) - N^L_z(-\tau) \right] / 2\tau
\label{Jz_avg}
\end{equation}
Similarly
\begin{equation}
\langle J_{tb} \rangle = \frac{1}{2\tau} \int_{-\tau}^{\tau}{dt} J_{tb}
=
-\left[ N^L_{tb}(\tau) - N^L_{tb}(-\tau) \right] / 2\tau
\label{Jtb_avg}
\end{equation}
The only difference between these two expressions lies in the
orbitals near the plane between atoms 0 and 1.
The integral defining $N_z^L$ contains part of the orbitals of
atom 0 and part of the orbitals of atom 1.
The integral defining $N_{tb}^L$ contains all of the orbitals
of atom 0 and none of the orbitals of atom 1. We can say
that the difference between $\langle J_{tb} \rangle$ and $\langle J_z \rangle$
must be bounded by
\begin{equation}
\left|
\sum_{a,b=0}^1  \;
i\hbar\sum_{i,j} \left[ G^<_{i_a,j_b}(\tau,\tau) -  G^<_{i_a,j_b}(-\tau,-\tau) \right]
\int d^3 r \phi^*_{j_b}(\rv) \phi_{i_a}(\rv)
\right| / 2\tau
\end{equation}
where $a$ and $b$ label the atom index and $i$ and $j$ are the orbital index.
If the quantity within the vertical brackets is finite, then the long
time average of the two current expressions must be identical.
We note that a rigorous mathematical proof requires careful
construction of the scattering states an example of which is
given in Appendix A of \cite{WingreenJacobsenWilkins2}, and also care
in taking the appropriate long
time limits for both the scattering state and the
adiabatic turn on of the device-contact coupling.

To conclude this section, we re-state the fact that the ``standard''
current expressions are the correct ones since they result in a
unitary transmission probability within the band of a periodic
system. The ``standard'' current expressions Eqs.
(\ref{JL_MeirWingreen}) and (\ref{J_general}) are identical in form
to the current expressions derived for an orthonormal basis. The
only difference is that the Hamiltonian matrix element ${\bf t}$ is
replaced everywhere by the effective Hamiltonian matrix element
$\btt = {\bf t} - E{\bf S}$.

Even with an explicit basis, one should
derive the current expression from
conservation laws that are consistent
with the Hamiltonian and basis such as
Eqs. (\ref{N_D}) - (\ref{continuity_t}).
A direct evaluation of the real space current operator
in an incomplete basis
will lead to a breakdown of the particle conservation
law and qualitatively incorrect results.
For a complete basis, the 2 approaches should
give the same result.

\section{Conclusion}
In summary, the use of the NEGF approach to quantum
electron transport developed in the early 1970s
is continuing to increase.
So much so that one could argue that it is
the most heavily used approach for modeling
quantum electron transport through nanostructures ranging from
semiconductors to carbon naontubes to molecules.

Currently, the heaviest usage is in the area of
molecular electronics in which the NEGF approach
is used to model current-voltage characteristics of
molecules.
All of the NEGF applications
in this area,
of which we are aware,
model coherent transport through the molecule.
The majority of the reported implementations
combine DFT with NEGF often integrating the NEGF
algorithm with existing ab-initio
commercial quantum chemistry or
computational materials codes.
These codes use an explicit non-orthogonal
localized orbital basis.
The formulation of the standard NEGF approach
to open system boundaries in a non-orthogonal basis
raises questions that have not been addressed in
the literature.

To address those questions,
we have re-derived
the standard NEGF theory in a non-orthogonal basis using the
second quantized formalism that underlies the theory.
This has allowed us to explore some of the approximations
that are commonly made, but of which, one finds little discussion.
The most fundamental approximation lies in the compatibility
of NEGF based on adiabatic perturbation theory and unitary
evolution of states with changes in the basis states and
corresponding Hilbert space.
This issue does not arise if one is only interested in $G^R$
since NEGF theory is not required to get $G^R$. It can be obtained
from the Heisenberg equation of motion and matrix algebra as shown
by the derivation of Eq. (\ref{GR_device_eom}).
In nonequilibrium, one must obtain $G^<$ from the contour
ordered Dyson equation, which, in principle, does not appear
to be compatible with perturbations in the basis states.
However, once $\Sigma^{RB}$ is fixed by the derivation of $G^R$,
the options for $\Sigma^{<B}$ and thus $G^<$ become limited,
and the standard expression for $\Sigma^{<B}$ Eqs.
(\ref{Sigma_<L}) and (\ref{Sigma_<R}) found
in the literature appears reasonable.

The second approximation underlies the derivation of
the ``standard''
current expression Eq. (\ref{JL_MeirWingreen}-\ref{FisherLee}).
We were unable to find a form for the local electron density
from which one could derive these expressions from a local continuity
equation.
Instead, we wrote a continuity equation for the total number
of electrons in a set of orbitals that define the ``device''
or left contact
rather than in a specific region
of space.
In the coherent limit, the resulting expression
for the transmission probability gives the correct
unitary value for a periodic system.

A direct evaluation of the real space current operator
in an incomplete basis
will lead to a breakdown of the particle conservation
law and qualitatively incorrect results.
For a complete basis, the standard approach and the
direct evaluation of the real space current operator should
give the same result.

%\noindent
\section*{Acknowledgments}
We acknowledge very helpful discussions with A. Korotkov.
This work was supported by the NSF (DMR-0103248),
DOD/DARPA/DMEA (DMEA90-02-2-0216),
and the Microelectronics Advanced Research Corporation
Focus Center on Nano Materials.

\newpage
%\bibliography{BIBLIOGRAPHY}

\begin{thebibliography}{157}
\bibitem[{non()}]{non-ortho_Wickes_thm}
\bibinfo{note}{We note in passing that Wicke's theorem applies as usual to the
  operators representing a non-orthogonal basis. This can be seen by performing
  the expansion using the field operators $\psi^{\dagger}(\rv)$ and $\psi(\rv)$
  and then integrating over $\rv$.}
\bibitem[{con()}]{continuum_approx}
\bibinfo{note}{{\mbox This assumes that the orbitals associated with atom 0 are
  in equilibrium. This may no longer be the case since the contact is now
  defined as a region of space rather than a set of orbitals.}}
\end{thebibliography}

\newpage
\noindent
{\bf Figure Captions}
\\

\noindent
1. Example calculation including incoherent scattering from acoustic phonons, polar optical phonons and interface roughness.
The simulations are compared against experimental data.
Reprinted from \cite{DRC96_proceedings}. Copyright 1996 IEEE.
\\

\noindent
2. Test matrix of In$_{0.47}$Ga$_{0.53}$As / AlAs RTDs.
The collector barrier is increased by one monolayer for the 3 RTDs from left to right. Forward and reverse bias current voltage curves are shown and overlaid on experimental data.
Reprinted from \cite{DRC97_proceedings}. Copyright 1997 IEEE.
\\

\noindent
3. Test matrix of In$_{0.47}$Ga$_{0.53}$As / In$_{0.48}$Al$_{0.52}$As RTDs.
(a-c) The well thickness is increased. (c-f) the undoped spacer layer is increased.
Reprinted from \cite{DRC97_proceedings}. Copyright 1997 IEEE.
\\

\noindent
4. Experimental data and simulation of the capacitance - voltage curve of
a MOS structure with a 3.1 nm oxide.  Reprinted from \cite{Lake_IEDM01}. Copyright 2001 IEEE.
\\

\noindent
5. Simulation of band profile of the Si / Si$_{0.5}$Ge$_{0.5}$ tunnel diode. Reprinted from \cite{Rommel_APL98}.
\\

\noindent
6.  Band profile of the delta-doped Si tunnel diode biased at 0.1V. Inset, 6 X valleys of the conduction band. Reprinted from \cite{CR_APL1}.
\\

\noindent
7. Three current components: TA phonon assisted, TO phonon assisted, and
direct or coherent current. Inset, 2D electron and hole dispersions. Reprinted from \cite{CR_APL1}.
\\

\noindent
8. Experimental I-V from ref. \cite{Thompson_APL99} corrected for
12 $\Omega$ series resistance overlaid on calculated
I-Vs with different energy broadening in the contacts as
shown in the legend. Reprinted from \cite{CR_JAP1}.
\\

\noindent
9. Comparison of the real and imaginary dispersion relations calculated
(a) from the parabolic single band model using the transverse conduction band mass of
0.19 m$_0$ and from the full-band model in the
transverse mass direction of the $X_4$ conduction band valley
and (b) from the single band model using the light hole mass of 0.16 m$_0$ and from
the full-band model in the valence band in the (001) direction.
The horizontal axis to the left of $0$ is imaginary $k$ and to the right of $0$ is
real $k$. Reprinted from \cite{CR_JAP1}.
\\

\noindent
10. Quantum cascade laser structure modeled in \cite{Lee_Wacker02}. Reprinted with permission from \cite{Lee_Wacker02}. Copyright 2002 American Physical Society.
\\

\noindent
11. Drain current and channel barrier height versus
extent of scattering region (starting from
source at -20 nm).
Inset: I$_D$ vs. V$_DS$ curves for V$_G = 0.6$ V. Reprinted with permission from \cite{Svizhenko_TED03}. Copyright 2003 IEEE.
\\

\noindent
12. Self-energy diagram for self-consistent Born approximation.
\\

\noindent
13. Keldysh time contour.

\newpage
\mbox{ }
\\
\\
\\
\\
\\
\mbox{ }
%\newpage
\mbox{ }
\\
\\
\\
\\
\\
\\
\\
\mbox{ }
%\newpage
\mbox{ }
\\
\\
\\
\\
\\
\\
\mbox{ }
%\newpage
\mbox{ }
\\
\\
\\
\\
\\
\\
\mbox{ }
%\newpage
\mbox{ }
\\
\\
\\
\\
\\
\\
\mbox{ }
%\newpage
\mbox{ }
\\
\\
\\
\\
\\
\\
\mbox{ }
%\newpage
\mbox{ }
\\
\\
\\
\\
\\
\mbox{ }
%\newpage
\mbox{ }
\\
\\
\\
\\
\\
\mbox{ }
%\newpage
\mbox{ }
\\
\\
\\
\\
\\
\mbox{ }
%\newpage
\mbox{ }
\\
\\
\\
\\
\\
\mbox{ }
%\newpage
\mbox{ }
\\
\\
\\
\\
\\
\\
\mbox{ }
%\newpage
\mbox{ }
\\
\\
\\
\\
\\
\mbox{ }
\newpage
\mbox{ }
\\
\\
\\
\\
\\
\begin{figure}[h]
\begin{center}
\resizebox{3.5in}{!}{ \includegraphics{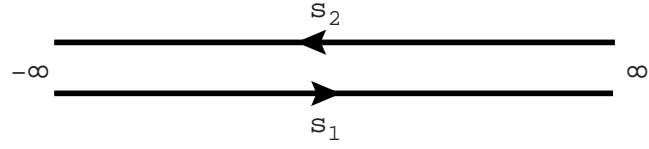} }
\end{center}
%\caption{\mbox{  }}
\caption{Keldysh time contour.}
\label{fig:Keldysh_contour}
\end{figure}
\mbox{ }
\newpage
%\mbox{ }
%\\
%\\
%\\
%\\
%\\
\end{document}